\documentclass{aa}  
\usepackage{graphicx}
\usepackage{txfonts}
\usepackage[breaklinks=true]{hyperref}
\usepackage{ctable}
\usepackage{mathtools}

\usepackage{soul}

\usepackage{natbib,twoopt}
\usepackage[breaklinks=true]{hyperref} 
\hypersetup{colorlinks=true, linkcolor=blue, urlcolor=blue, citecolor=blue}
\bibpunct{(}{)}{;}{a}{}{,} 
\makeatletter
\newcommandtwoopt{\citeads}[3][][]{\href{http://adsabs.harvard.edu/abs/#3}%
{\def\hyper@linkstart##1##2{}%
\let\hyper@linkend\@empty\citealp[#1][#2]{#3}}}
\newcommandtwoopt{\citepads}[3][][]{\href{http://adsabs.harvard.edu/abs/#3}%
{\def\hyper@linkstart##1##2{}%
\let\hyper@linkend\@empty\citep[#1][#2]{#3}}}
\newcommandtwoopt{\citetads}[3][][]{\href{http://adsabs.harvard.edu/abs/#3}%
{\def\hyper@linkstart##1##2{}%
\let\hyper@linkend\@empty\citet[#1][#2]{#3}}}
\newcommandtwoopt{\citeyearads}[3][][]%
{\href{http://adsabs.harvard.edu/abs/#3}
{\def\hyper@linkstart##1##2{}%
\let\hyper@linkend\@empty\citeyear[#1][#2]{#3}}}
\makeatother



\newcommand{\adeg}[1]{{#1}$^{\circ}$}
\newcommand{\amin}[1]{{#1}$^\prime$}
\newcommand{\asec}[1]{{#1}$^{\prime\prime}$}

\newcommand{\mjy}[1]{{#1} mJy}
\newcommand{\mjybeam}[1]{{#1} mJy beam$^{-1}$}

\newcommand{\ujybeam}[1]{{#1} $\mu$Jy beam$^{-1}$}

\newcommand{\bootes}{Bo\"otes}

\definecolor{mygreen}{rgb}{0.0, 0.42, 0.0}

\usepackage[mathlines,switch]{lineno}

\newcommand{\napt}{8994~}
\newcommand{\nhba}{36767~}
\newcommand{\nlba}{1948~}
\newcommand{\naptrec}{8742~}
\newcommand{\nhbarec}{8153~}
\newcommand{\nlbarec}{1675~}

\usepackage{lipsum}

\newcommand\blfootnote[1]{%
  \begingroup
  \renewcommand\thefootnote{}\footnote{#1}%
  \addtocounter{footnote}{-1}%
  \endgroup
}

\begin{document} 
    \title{Apertif 1.4\,GHz continuum observations of the \bootes\ field 
           and their combined view with LOFAR}
   \author{
            A.\,M.\,Kutkin
            \inst{1^*}
        \and
            T.\,A.\,Oosterloo
            \inst{1,2}
        \and
            R.\,Morganti
            \inst{1,2}
        \and
            A.\,R.\,Offringa
            \inst{1,2}
        \and
            E.\,A.\,K.\,Adams
            \inst{1,2}
        \and
            B.\,Adebahr
            \inst{3}
        \and
            H.\,D\'{e}nes
            \inst{4,1}
        \and
            K.\,M.\,Hess
            \inst{1,5,6}
        \and
            J.\,M.\,van\,der\,Hulst
            \inst{2}
        \and
            W.\,J.\,G.\,de\,Blok
            \inst{1,7,2}
        \and
            A.\,Bozkurt
            \inst{1}
        \and
            W.\,A.\,van\,Cappellen
            \inst{1}
        \and
            A.\,W.\,Gunst
            \inst{1}
        \and
            H.\,A.\,Holties
            \inst{1}
        \and
            J.\,van\,Leeuwen
            \inst{1}
        \and
            G.\,M.\,Loose
            \inst{1}
        \and
            L.\,C.\,Oostrum
            \inst{1,8,9}
        \and
            D.\,Vohl
            \inst{1}
        \and
            S.\,J.\,Wijnholds
            \inst{1}
        \and
            J.\,Ziemke
            \inst{1,10}
        }
          
    \institute{
        ASTRON, The Netherlands Institute for Radio Astronomy, Oude Hoogeveensedijk 4, 7991 PD, Dwingeloo, The Netherlands\\
        \email{kutkin@astron.nl}
    \and
        Kapteyn Astronomical Institute, P.O. Box 800, 9700 AV Groningen, The Netherlands
    \and
        Astronomisches Institut der Ruhr-Universit{\"a}t Bochum (AIRUB), Universit{\"a}tsstrasse 150, 44780 Bochum, Germany
    \and
        School of Physical Sciences and Nanotechnology, Yachay Tech University, Hacienda San Jos\'{e} S/N, 100119, Urcuqu\'{i}, Ecuador
    \and
        Instituto de Astrof\'{i}sica de Andaluc\'{i}a (CSIC), Glorieta de la Astronom\'{i}a s/n, 18008 Granada, Spain
    \and 
        Department of Space, Earth and Environment, Chalmers University of
        Technology, Onsala Space Observatory, 43992 Onsala, Sweden
    \and
         Dept.\ of Astronomy, Univ.\ of Cape Town, Private Bag X3, Rondebosch 7701, South Africa        
     \and
     Anton Pannekoek Institute, University of Amsterdam, Postbus 94249, 1090 GE Amsterdam, The Netherlands
    \and
        Netherlands eScience Center, Science Park 140, 1098 XG, Amsterdam, The Netherlands
    \and
        University of Oslo Center for Information Technology, P.O. Box 1059, 0316 Oslo, Norway
    }

 
  \abstract
  {
We present a new image of a 26.5 square degree region in the \bootes\ constellation obtained at 1.4~GHz using the Aperture Tile in Focus (Apertif) system on the Westerbork Synthesis Radio Telescope. We use a newly developed processing pipeline which includes direction-dependent self-calibration which provides a significant improvement of the quality of the images compared to those released as part of the Apertif first data release. 
For the  \bootes\ region, we mosaic 187 Apertif images and extract a source catalog. The mosaic image has an angular resolution of \asec{$27\times11.5$} and a median background noise of \ujybeam{40}. The catalog has \napt sources and is complete down to the \mjy{0.3} level. 
We combine the Apertif image with LOFAR images of the \bootes\ field at 54 and 150 MHz to study spectral properties of the sources.
We find a spectral flattening towards low flux density sources. Using the spectral index limits from Apertif non-detections we derive that up to 9\% of the sources have ultra-steep spectra with a slope steeper than --1.2. 
Steepening of the spectral index with increasing redshift is also seen in the data showing a different dependency for the low-frequency spectral index and the high frequency one. This can be explained by a population of sources having concave radio spectra with a turnover frequency around the LOFAR band. Additionally, we discuss cases of individual extended sources with an interesting resolved spectral structure.
With the improved pipeline, we aim to continue processing data from the Apertif wide-area surveys and release the improved 1.4-GHz images of several famous fields.
}


\maketitle \blfootnote{\noindent Tables 1 and 3 are only available in electronic form at \url{http://vo.astron.nl} and the CDS via anonymous ftp to cdsarc.u-strasbg.fr (130.79.128.5) or via \url{http://cdsarc.u-strasbg.fr}}
\blfootnote{\noindent *kutkin@astron.nl}


\section{Introduction}\label{sec:intro}

Apertif (APERture Tile In Focus) is a phased array feed (PAF) system for the Westerbork synthesis radio telescope (WSRT) designed for performing wide field surveys of the northern sky~\citep{2022A&A...658A.146V}. During its operations in 2019 – 2022, the Apertif surveys covered about 2300 square degrees (Hess et al. in prep). The first imaging data release took place in 2019 and is described in~\cite{2022A&A...667A..38A}.

One of the added values of the Apertif surveys is their synergy with the LOFAR (Low Frequency Array, \citealt{2013A&A...556A...2V}) surveys. LOFAR provides radio images with resolution of \asec{6} at 150\,MHz (High Band Antenna, HBA; \citealt{2022A&A...659A...1S}) and \asec{15} resolution at 50~MHz (Low Band Antenna, LBA; \citealt{deGasperin21, 2023arXiv230112724D}), the latter having a resolution particularly close to the one of the Apertif images. 
Having comparable angular resolution and sensitivity (for spectral indices typical of extragalactic radio sources), the combination of these three surveys provides information about radio spectra of the sources at MHz--GHz frequencies. 
Some examples of results from such combinations are illustrated in \cite{2021Galax...9...88M, 2021A&A...648A...9M}, \citet[][hereafter K22]{K22}, Shulevski et al.\,(in prep), Adebahr et al.\,(in prep). 

Famous fields observed in multiple bands by a variety of telescopes are ideal regions for probing the spectral properties of radio sources, since they also provide a wealth of ancillary data, most notably host galaxy properties and redshifts. One such field for which the study of radio spectra can be expanded is the \bootes~field (see description of this field and the available ancillary data by \citealt{2021A&A...655A..40W}). Considering the availability of existing deep LOFAR HBA and LBA observations, it is a prime target to be imaged at 1.4 GHz by Apertif in order to allow the analysis of the sources using three frequencies simultaneously.

Since the start of the study of radio sources, radio spectra have been recognized as a powerful tool for deriving important properties of their radio emission, such as the injection mechanism and the presence of absorption, and for tracing their history and evolution \citep{1962SvA.....6..317K, 1968ARA&A...6..321S, 1970ranp.book.....P, 1991ApJ...383..554C, 1992ARA&A..30..575C,2017NatAs...1..596M}.
Although the spectral shape of radio sources can be, to first order, approximated by a power law ($S\propto\nu^\alpha$, where $\alpha$ is the spectral index), deviations from this simple form provide key signatures for tracing physical and evolutionary properties of a radio source. 
For example, inverted spectral indices observed at low frequencies (i.e. a spectral index higher than that of the higher frequency part of the spectrum) can indicate the presence of an absorption mechanism, such as free-free absorption or of synchrotron self-absorption, which might lead to a peaked spectrum shape. An important group of sources that can be identified in this way is that of young radio galaxies with newly born jets~\citep[see][for a review]{2021A&ARv..29....3O}.
At the other extreme -- for sources in which the central activity has stopped, or substantially decreased (i.e. dying radio sources), the spectrum is expected to show steeping, starting at higher frequencies, becoming \textit{ultra-steep} (USS) with $\alpha \lesssim -1.2$ ~\citep{1994A&A...285...27K, 2001AJ....122.1172S}. 
Identifying such objects helps to shed light on the evolution cycle of extragalactic radio sources. USS sources have been also used to pinpoints high-redshift radio galaxies because of the observed trend between spectral index and redshift~\cite[see e.g.,][]{1979A&A....80...13B, 1994A&AS..108...79R, 2006MNRAS.371..852K, 2018MNRAS.480.2726M}.
In addition, spectral index images reveal different electron populations within a source (hot spots, remnant lobes, tails) which give clues on the on-going process and  the evolution of radio sources~\cite[e.g.,][]{2017MNRAS.469..639H}.

Large and deep samples with observations covering multiple parts of the radio spectrum 
are an essential ingredient for identifying these extreme phases in the life of radio sources so that a better inventory can be made of the relevance of the various processes.
In this work we present a new deep continuum mosaic image and source catalog obtained at 1400 MHz with Apertif, covering the full area of the \bootes~field imaged earlier at 54 and 150 MHz by LOFAR (Williams et al. and Tasse et al. 2021). Compared to the images of the First Apertif Data Release, the present study makes use of an improved processing pipeline which now includes direction-dependent calibration, resulting in much better image quality. Based on these new images,  we analyze and discuss spectral properties of the sources. 

The description of the new pipeline is given in Sect.~\ref{sec:pipeline}. The new image and the catalog are described in Sect.~\ref{sec:apertif_image_catalog}. The LOFAR data and further processing are described in Sect.~\ref{sec:lofar_data}. The spectral indices are analyzed and discussed in Sect.\ \ref{sec:SI}.

\begin{figure*}
    \centering
    \includegraphics[width=2\columnwidth]{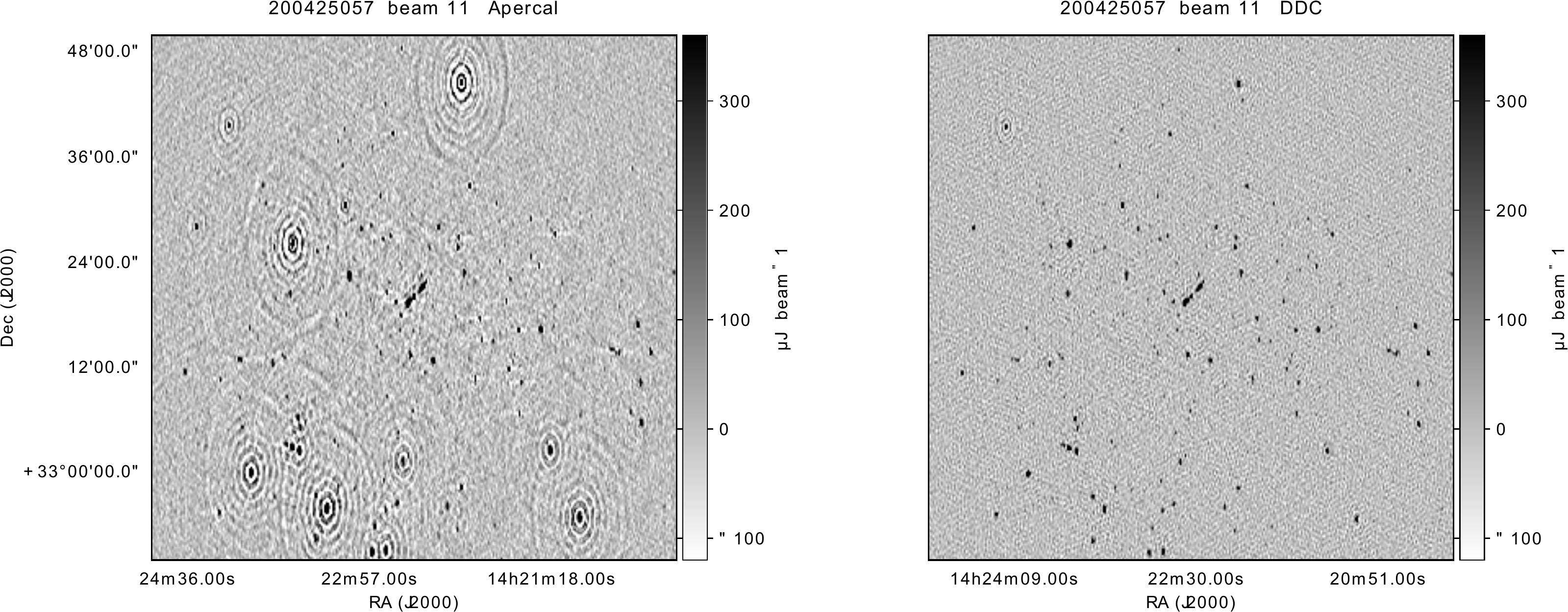}
    \caption{Example of the improvement in image quality  as the result of including direction-dependent effects in the calibration of Apertif data. {\sl Left panel:} Apertif image obtained with the old, direction-independent pipeline. {\sl Right panel:}  Image made from the same data, but applying the new, direction-dependent pipeline. 
            }
    \label{fig:apercal_vs_ddcal}
\end{figure*}

\begin{figure*}
    \centering
    \includegraphics[width=\linewidth]{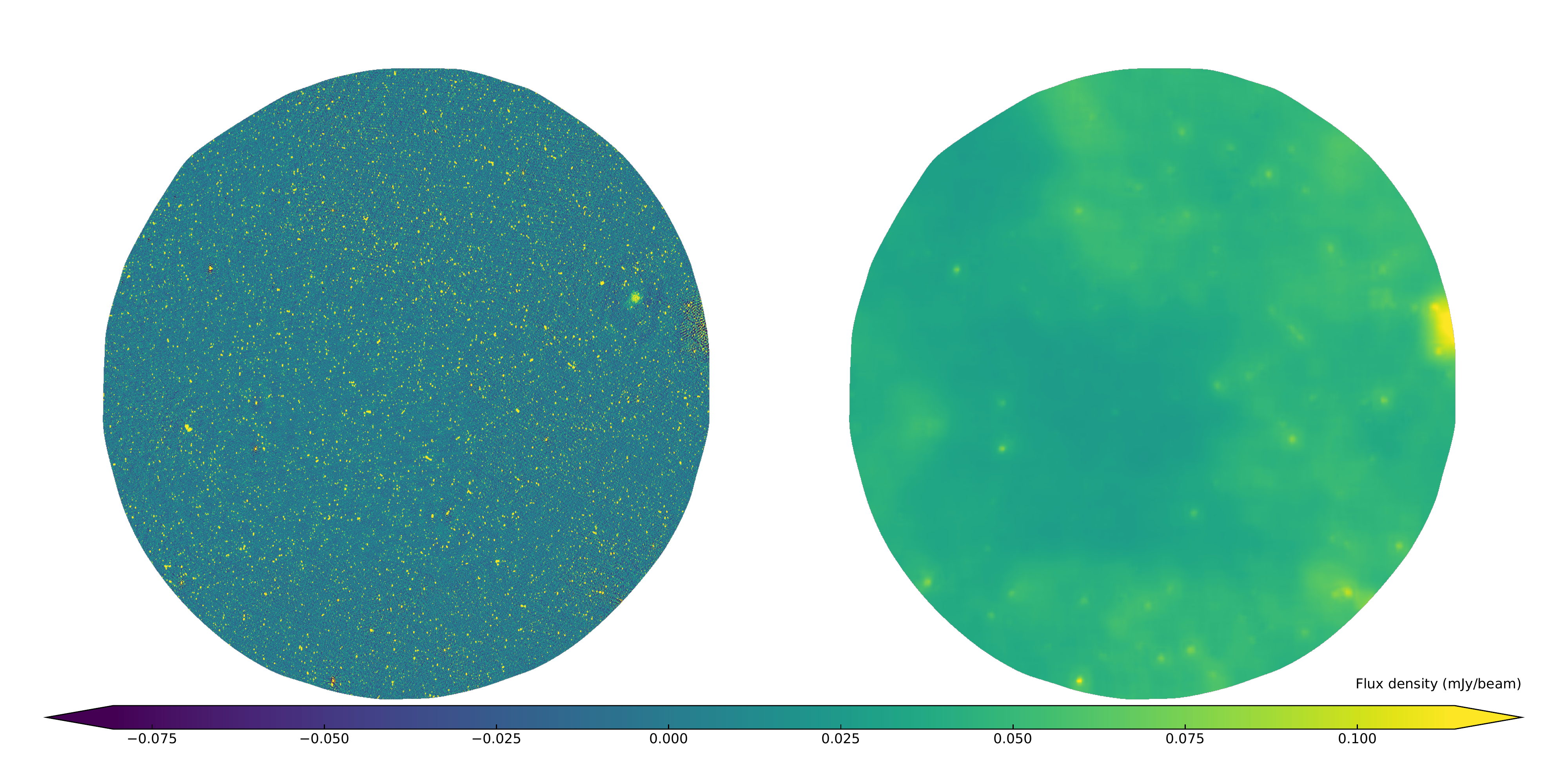}
    \caption{Apertif mosaic of the \bootes\ region and the noise RMS map. The sky coverage matches that of the LOFAR HBA deep field image. The image is centered at RA = 14$^{\rm h}$31$^{\rm m}$53$^{\rm s}$, DEC = +34$^\circ$27$^\prime$43$^{\prime\prime}$ and covers an area of 26.5 square degrees.}
    \label{fig:mosaic}
\end{figure*}


\section{Processing pipeline}\label{sec:pipeline}

The default Apertif pipeline, \texttt{Apercal}\footnote{https://github.com/apertif/apercal}, includes automatic calibration and imaging procedures~\citep{2022A&C....3800514A}. However, it provides only a direction-independent calibration which, in the case of Apertif, is often not sufficient for obtaining science-quality images. The resulting images often suffer from direction-dependent effects (DDEs) arising from malfunctioning PAF elements. Faulty PAF elements effectively lead to large errors in the shape of a compound beam of a given antenna. In the case of WSRT, these effects manifest themselves as concentric elliptical artifacts around sources (see Fig.\ \ref{fig:apercal_vs_ddcal} for an example). These artifacts significantly reduce the dynamic range of the images and complicate the source finding procedure. Overall, more than half of all Apertif images are affected by DDEs. 

We have developed and applied a new procedure which includes both direction-independent (DI) and direction-dependent (DD) self-calibration to process Apertif data. It is based on the LOFAR Default Preprocessing Pipeline~\citep[DPPP/DP3;][]{DP3} and the \texttt{wsclean} imaging package~\citep{wsclean}. The code and documentation of the new pipeline are available at GitLab\footnote{https://git.astron.nl/kutkin/apipeline}.

The image of each compound beam of an observations is calibrated independently. The new pipeline starts with the preflagged cross-calibrated continuum visibilities produced with the initial steps of \texttt{Apercal}. 
For many of these data sets, additional flagging is done of antenna-beam combinations for which the direction-dependent errors are very large.
Following this, a DI calibration is performed. This step includes three cycles of self-calibration and imaging. First, a phase-only self-calibration with a 10 minute solution interval is run, followed by a phase-only calibration with 30 seconds solution interval. The final step in the DI calibration is an amplitude and phase calibration with a long solution interval of 1 hour. The second and third \texttt{CLEAN} step in the DI self calibration are done using masks based on the local signal-to-noise estimated from the residual images of the previous step. 
The number of sub-steps, type of the calibration, parameters for creating the local noise images, and the solution intervals used were determined before running the pipeline on all images  by manually experimenting with these parameters and analyzing the results through the validation procedure described in~\cite{2022A&A...667A..38A}. This resulted in a final set of parameters used for all images.

After the DI calibration, a clustering procedure is performed. The final \texttt{CLEAN} model obtained after the DI calibration is segmented using Voronoi tessellation with cluster centers located at the ten brightest sources. Using this segmented model,  DD calibration is performed by calibrating each segment independently. This step is performed using DP3 \texttt{ddecal} with the parameter \texttt{subtract=True}, meaning that the visibilities are subtracted from the DI calibrated data with the DD calibration solutions applied. The residual visibilities, free from DDEs, are then imaged and the final DI model is restored on this image.
This final compound beam image is produced with a size of 3072$\times$3072 pixels and pixel scale of \asec{3}.

To illustrate the quality improvement over images produced with the original DI pipeline, Fig.\ \ref{fig:apercal_vs_ddcal} shows an image obtained with this original DI pipeline as well as the image made from the same data, but using the new DD pipeline. As can be seen, the elliptical artifacts due to DDEs present in the image produced with the original pipeline have been corrected for and the quality of the images has significantly improved. The flux density of the sources does not differ by more than a few percent between the images. 


\section{Apertif image and catalog}
\label{sec:apertif_image_catalog}

\subsection{Mosaic image}
In this work we focus on the sky area of approximately $30$ square degrees of the Bo\"otes field as observed with LOFAR (see Sect.~\ref{sec:lofar_data}). This area has been covered by Apertif with 187 compound beam images from 8 different survey observations performed between April 2019 and November 2021. The self-calibration and imaging procedure was performed with the newly developed pipeline described in Sect.~\ref{sec:pipeline}. The direction-dependent calibration allowed us to achieve a significant improvement on the quality of the individual images (as shown in Figure~\ref{fig:apercal_vs_ddcal}). This is quantified by the fraction of compound beam images that pass the Apertif quality validation (see \citealt{2022A&A...667A..38A} for details) which increased from 40\% using the default pipeline to 98\% with the new one.

To obtain the final mosaic image, first the astrometric accuracy of the images was improved by cross-matching sources from each individual image with those in the LOFAR HBA image (labeled M2 and described in Sect.~\ref{sec:lofar_data}) and correcting the astrometry of the Apertif images with the median position offset. The typical offset was found to be of the order of \asec{1} with only a few exceptions. 
After this, the images were combined in the standard way using the \texttt{amosaic}\footnote{https://github.com/akutkin/amosaic} mosaic package  described in K22. This procedure includes a correction of the images for the compound beam shapes (in this work we used the same beam models as in K22), re-convolving the images to a common angular resolution, and re-projecting them onto the common sky grid. Because individual Apertif images overlap, the mosaicking leads to an increase in sensitivity. The mosaic image is shown in the left panel of Fig.~\ref{fig:mosaic}, and the noise map is shown in the right panel of this figure. The angular resolution of the mosaic image is $27^{\prime\prime}\times$ 11\farcs5. The local noise varies over the image between \ujybeam{25} and \ujybeam{50} with a median value of \ujybeam{40}. The image is available for download at \url{https://vo.astron.nl}.

\subsection{The catalog}
\label{sec:catalog}
We extracted the Apertif source catalog (C1) using the Python Blob Detector and Source Finder \citep[PyBDSF;][]{2015ascl.soft02007M}.
Similar to~\cite{2019A&A...622A...1S}, we used a peak and island detection threshold of 5 $\sigma$ and 4 $\sigma$ respectively, and the size of a sliding box to estimate the local RMS was set to $30\times30\times b_{\rm maj}$, where $b_{\rm maj}$ is the major axis of the synthesized beam. This size was set to automatically decrease to $12\times12\times b_{\rm maj}$ near bright sources to capture noise variations more accurately. 
For consistency, the same PyBDSF settings were used  for the smoothed images described below in Sect.~\ref{sec:lofar_data}. PyBDSF only provides  fitting errors for the flux density and position measurements, and gives incorrect values for the errors in deconvolved shape parameters. Those errors were adjusted and re-calculated in the same way as described in K22 (Appendix A). 

An example of the catalog structure is shown in Table~\ref{tab:catalog}. The columns designations are: 
(1): Apertif source name; 
(2,4): RA and Dec; 
(3,5): RA and Dec errors; 
(6,8): total and peak flux density; 
(7,9): integrated and peak flux density uncertainties;
(10,12,14): Deconvolved major and minor source size and position angle. A source size value 0.0 means that the PSF cannot be deconvolved from the fitted source along the given axis, and the corresponding uncertainty, represents an upper size limit estimate. The position angle is given in degrees ranging from --90$^\circ$ to 90$^\circ$ west to east through north. When both major and minor sizes are given as 0.0, the position angle is omitted; 
(11,13,15): uncertainties of the major and minor source size and position angle;
(16): local background noise rms; 
(17): Source type as classified by PyBDSF, with 'S' as an isolated source fitted with a single Gaussian; 'C' as sources that were fit by a single Gaussian but are within an island of emission that also contains other sources, and 'M' as sources fitted with multiple Gaussians. 
The resulting catalog contains \napt sources, among which 8\% were modeled with multiple Gaussians indicating a complex structure (\texttt{S\_Code=M}), and almost all other sources were modeled with a single Gaussian (\texttt{S\_Code=S}). 

After compiling the catalog, we cross-matched it with the NVSS~\citep{1998AJ....115.1693C} to  compared the flux density scale. The median total flux density ratio is 0.98, providing an additional check on the calibration, mosaicing and cataloging procedures. 

\begin{table*}
    {\scriptsize
    \caption{Apertif source catalog. \label{tab:catalog}}
   {\setlength{\tabcolsep}{5pt}
    \begin{tabular}{lcccccccccccccccc}
    \hline\hline
    Name & RA & $\sigma_\mathrm{RA}$ & Dec & $\sigma_\mathrm{Dec}$ & $S_\mathrm{total}$ &  $\sigma_{S_\mathrm{total}}$ &  $S_\mathrm{peak}$ &  $\sigma_{S_\mathrm{peak}}$ &  Maj &  $\sigma_\mathrm{Maj}$ & Min &  $\sigma_\mathrm{Min}$ & PA & $\sigma_\mathrm{PA}$ & RMS & S\_Code \\
    ~  & [\adeg{}] & [\asec{}] & [\adeg{}] & [\asec{}] & mJy & mJy & mJy/bm & mJy/bm & [\asec{}] & [\asec{}] & [\asec{}] & [\asec{}] & [\adeg{}] & [\adeg{}] & $\muup$Jy/bm &  (S/M/C) \\
    (1) & (2) & (3) & (4) & (5) & (6) & (7) & (8) & (9) & (10) & (11) & (12) & (13) & (14) & (15) & (16) & (17) \\
     \hline
     APTF\_J142347+360215 & 215.9489 & 1.7 & 36.0378 & 1.0 & 5.28 & 0.37 & 1.68 & 0.10 & 78.1 &  3.6 &  0.0 & 2.8 & --18.3 &   1.5 & 51.2 & M    \\
    APTF\_J142347+334237 & 215.9491 & 1.7 & 33.7103 & 2.5 & 0.54 & 0.11 & 0.40 & 0.05 & 14.8 &  7.6 &  6.3 & 5.3 &  --1.8 &  16.6 & 49.0 & S    \\
    APTF\_J142348+333741 & 215.9505 & 1.7 & 33.6281 & 2.3 & 0.42 & 0.09 & 0.39 & 0.05 &  0.0 & 16.1 &  0.0 & 6.9 & \dots & \dots & 50.0 & S    \\
    APTF\_J142348+325706 & 215.9506 & 1.1 & 32.9519 & 1.1 & 1.67 & 0.13 & 1.53 & 0.09 &  0.0 &  7.8 &  0.0 & 5.4 & \dots & \dots & 53.3 & S    \\
    APTF\_J142348+360527 & 215.9512 & 1.4 & 36.0911 & 1.4 & 8.50 & 0.53 & 1.81 & 0.10 & 63.6 &  3.7 & 12.0 & 3.0 & --53.1 &   3.7 & 51.0 & M    \\
    APTF\_J142348+345034 & 215.9513 & 1.0 & 34.8429 & 0.9 & 2.03 & 0.13 & 1.71 & 0.10 &  9.0 &  1.4 &  0.0 & 3.3 & --57.9 &  13.1 & 43.6 & S    \\
    APTF\_J142348+331951 & 215.9513 & 1.5 & 33.3309 & 2.0 & 0.46 & 0.09 & 0.45 & 0.05 &  0.0 & 12.6 &  0.0 & 3.2 & \dots & \dots & 47.3 & S    \\
    APTF\_J142348+345502 & 215.9514 & 1.1 & 34.9174 & 1.0 & 1.46 & 0.11 & 1.33 & 0.08 &  6.7 &  3.2 &  3.5 & 1.4 &  10.6 &  12.6 & 43.9 & S    \\
    APTF\_J142348+324234 & 215.9519 & 1.8 & 32.7095 & 2.5 & 0.51 & 0.12 & 0.43 & 0.06 &  7.6 & 10.0 &  4.6 & 3.0 &  53.7 &  75.9 & 59.3 & S    \\
    APTF\_J142348+364920 & 215.9529 & 1.1 & 36.8223 & 1.1 & 2.25 & 0.16 & 1.60 & 0.09 & 19.0 &  2.2 &  4.4 & 1.3 &  11.2 &   3.9 & 52.2 & S    \\
    APTF\_J142348+323548 & 215.9529 & 1.1 & 32.5968 & 1.1 & 1.96 & 0.16 & 1.69 & 0.11 &  8.5 &  2.6 &  0.0 & 5.5 &  58.2 &  26.6 & 67.1 & S    \\
    APTF\_J142348+332547 & 215.9538 & 1.3 & 33.4297 & 1.5 & 0.91 & 0.11 & 0.80 & 0.06 &  8.2 &  5.0 &  1.2 & 5.2 &  41.8 &  37.8 & 51.3 & S    \\
    APTF\_J142348+321540 & 215.9540 & 1.1 & 32.2613 & 1.1 & 1.50 & 0.12 & 1.44 & 0.09 &  5.2 &  4.1 &  0.0 & 3.6 & --27.3 &  29.1 & 56.2 & S    \\

    \hline
    \end{tabular}
    }
    \tablefoot{Sample of the catalog records. Descriptions of the columns are given in Sect.~\ref{sec:apertif_image_catalog}. The full table containing \napt entries will be available in machine-readable format at \url{https://vo.astron.nl} and through CDS.}
    }
\end{table*}


\begin{figure}
    \centering
    \includegraphics[width=\columnwidth]{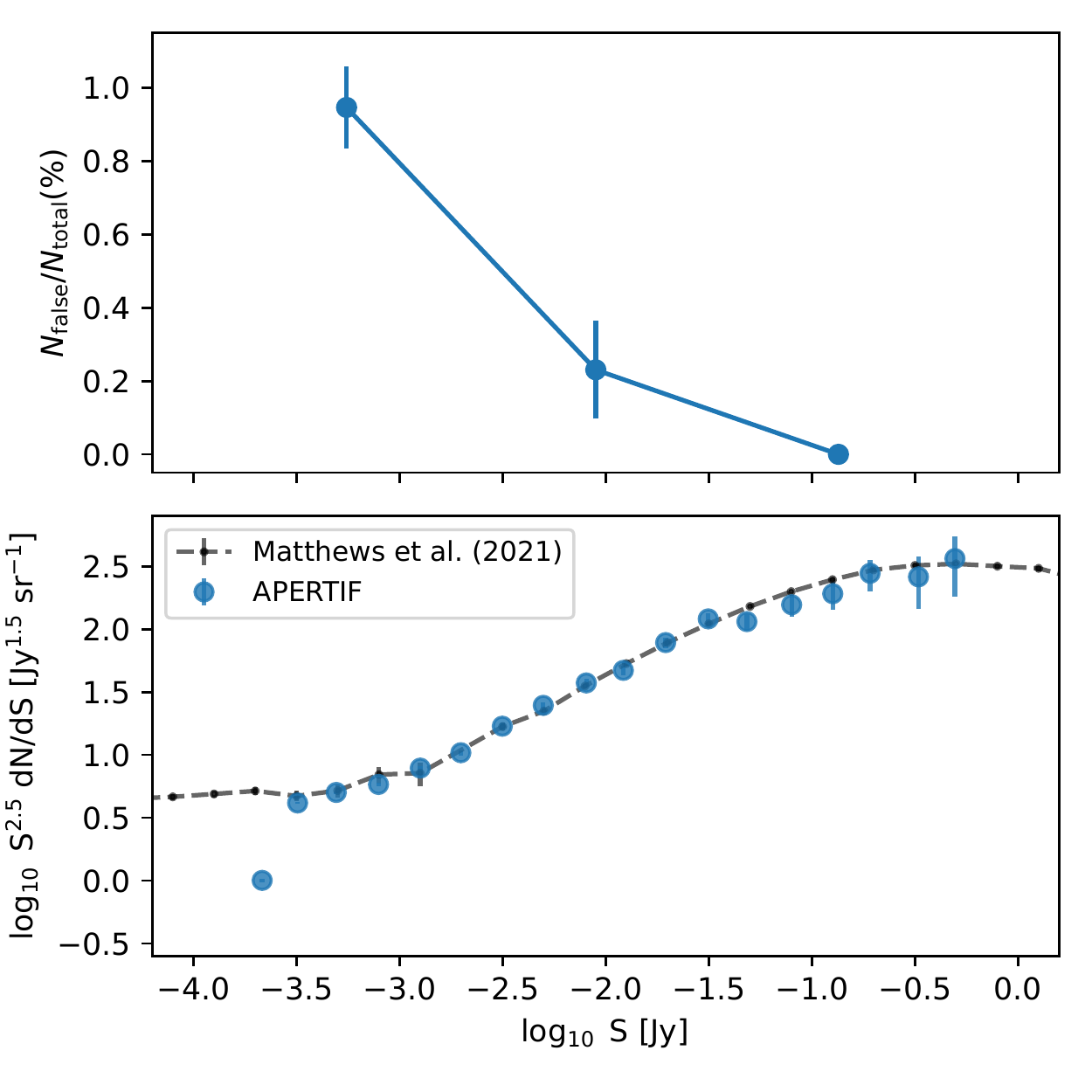}
    \caption{False detections fraction and differential source counts in the Apertif catalog.}
    \label{fig:counts}
\end{figure}


\begin{figure*}
    \centering
    \includegraphics[width=\linewidth]{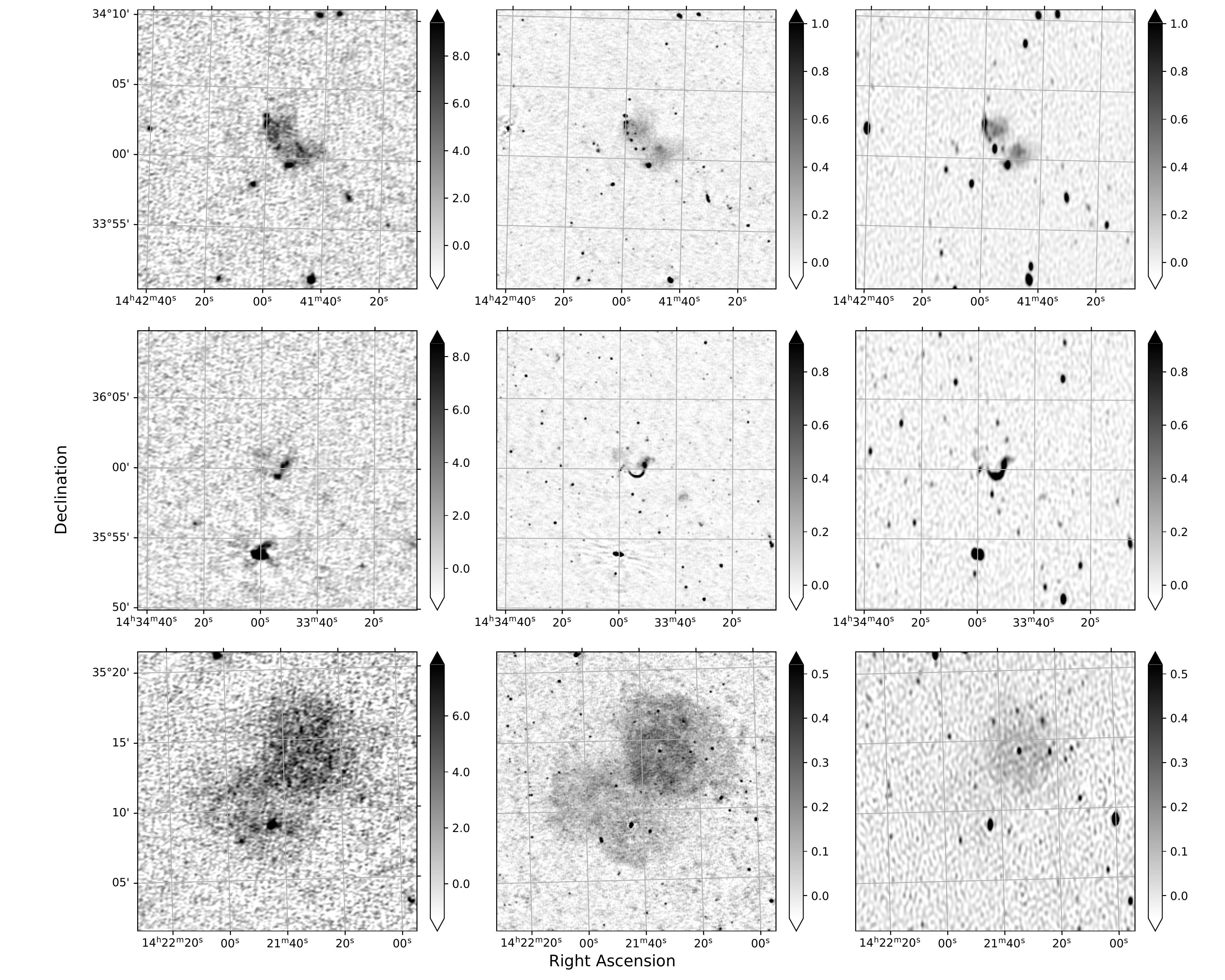}
    \caption{Three images as seen by LBA (left column), HBA (middle column) and Apertif (right column). The color-bar scale is shown in \mjybeam{}. The object shown in the bottom row has been studied by~\cite{2014MNRAS.440.1542D} and a new, detailed study using also the Apertif data will be presented by Adebahr et al. (in prep).
    }
    \label{fig:zooms}
\end{figure*}


\subsection{Reliability and completeness}
\label{sec:completeness}
We estimated the reliability and completeness of the catalog in the same way as described in K22 (see their Sect.\ 5.5). The former was obtained from PyBDSF false detections in the inverted image within an assumption of a symmetric noise distribution. The number of false detections in the Apertif catalog does not exceed 1.5 percent over the full flux density range. The completeness was estimated by comparing the differential source counts corrected for the false positive detection rate to the results published by \citet{2021ApJ...909..193M}, who provided  comprehensive source counts over a wide flux density range at 1.4\,GHz. Both false detection distribution and differential source counts as a function of total flux density are shown in Figure~\ref{fig:counts}. It is seen that the catalog remains complete down to the level of about \mjy{0.3}. Note, that this level is calculated for the entire image, and a ``local'' completeness can be better as the local noise varies significantly over the mosaic (see Figure~\ref{fig:mosaic}). 


\section{LOFAR data and cross-matching of the catalogs}
\label{sec:lofar_data}

We used the publicly available LOFAR HBA and LBA mosaic images (referred to as M2, M3 in the remaining of the paper and Table \ref{tab:data}) and the corresponding catalogs (C2, C3) of the Bo\"otes area~\citep{2021A&A...648A...1T, 2021A&A...655A..40W}\footnote{\url{https://lofar-surveys.org/}}. 
Examples of  sources as seen by LOFAR and Apertif are shown in Fig.~\ref{fig:zooms}. 

Different angular resolution complicates cross-matching of the catalogs, leading to inaccurate spectral index measurements.
This is especially significant when a source is resolved in one of the images remaining unresolved in another. For the further analysis, we re-projected and smoothed the Apertif and LOFAR mosaic images to the same sky footprint and a common angular resolution of 27$\times$15\asec{}.

The images and the corresponding catalogs (see below for a description) are listed in Table~\ref{tab:data}. In column 5 of this table we report the median value of the background noise RMS derived by PyBDSF at the locations of the sources (\texttt{Isl\_rms}). It is seen that degrading resolution leads to an increase of the image noise, which has especially strong impact on the HBA image M2. However, for the further analysis, the benefits of the images at the same resolution are of higher importance. 

We cross-matched the catalogs C4, C5 and C6 using a \asec{5} matching radius which results in 1286 sources in common between all three catalogs, 5605 sources in common between C4 and C5 and 1401 sources in common between C5 and C6.

\begin{table}[]
    \scriptsize{
    \caption{Images, catalogs and their properties}
    \begin{tabular}{lccccc}
    \hline\hline
        Instrument & Frequency & Resolution & N sources & Noise & I/C \\
                   & [MHz]      & [\asec{}]  & & [\ujybeam{}] & \\
        ~~~~~~(1) & (2) & (3) & (4) & (5) & (6) \\
    \hline     
         Apertif & 1400 & 27$\times$11.5 & \napt & 40  & M1 / C1\\
         LOFAR-HBA & 150 & 6$\times$6 & \nhba & 50 & M2 / C2\\
         LOFAR-LBA & 54 & 15$\times$15 & \nlba & 1152 & M3 / C3\\
         Apertif & 1400 & 27$\times$15 & \naptrec & 42 & M4 / C4\\
         LOFAR-HBA & 150 & 27$\times$15 & \nhbarec & 185 & M5 / C5\\
         LOFAR-LBA & 54 & 27$\times$15 & \nlbarec & 1320  & M6 / C6\\
    \hline
    \end{tabular}
    \tablefoot{Columns designation: (1): Instrument/Survey (2): Approximate central frequency; (3): Image resolution; (4): Number of sources in the catalog; (5): Median value of the noise RMS reported by PyBDSF;  (6): Mosaic image and catalog reference in the text;
    }
    \label{tab:data}
    }
\end{table}
 
\begin{table*}
\centering
{\scriptsize
\caption{Spectral indices of the sources \label{tab:tab_si}}
\begin{tabular}{cccccccccccccc}
\hline\hline
        RA$_{\mathrm A}$ & DEC$_{\mathrm A}$ & $S_\mathrm{A,int}$ & RA$_{\mathrm H}$ & DEC$_{\mathrm H}$ & $S_\mathrm{H,int}$ & RA$_{\mathrm L}$ & DEC$_{\mathrm L}$ & $S_\mathrm{L,int}$ & $\alpha_{\mathrm high}$ & $\sigma_{\alpha_{\mathrm high}}$ & $\alpha_{\mathrm low}$ & $\sigma_{\alpha_{\mathrm low}}$ & z \\
        $[^{\circ}]$ & $[^{\circ}]$ & mJy & $[^{\circ}]$ & $[^{\circ}]$ & [mJy] & $[^{\circ}]$ & $[^{\circ}]$ & [mJy] \\
        (1) & (2) & (3) & (4) & (5) & (6) & (7) & (8) & (9) & (10) & (11) & (12) & (13) & (14)\\        
\hline
    215.7532 & 34.9463 &   0.3 &    \dots &   \dots & \dots &    \dots &   \dots & \dots & >--0.63 & \dots &   \dots & \dots & \dots \\
    215.9791 & 33.4191 &   0.4 & 215.9795 & 33.4185 &   2.0 &    \dots &   \dots & \dots & -0.743 & 0.164 & >--1.32 & \dots & \dots \\
    218.1798 & 32.3572 &   7.8 &    \dots &   \dots & \dots &    \dots &   \dots & \dots &  >0.83 & \dots &   \dots & \dots & \dots \\
    218.2465 & 32.6096 &   0.3 & 218.2457 & 32.6089 &   2.2 &    \dots &   \dots & \dots & --0.947 & 0.207 & >--1.07 & \dots & \dots \\
       \dots &   \dots & \dots & 217.0903 & 34.4777 &   1.3 &    \dots &   \dots & \dots & <--0.74 & \dots & >--1.33 & \dots & 0.160 \\
    217.4835 & 35.3232 &   1.7 & 217.4825 & 35.3238 &  12.0 & 217.4830 & 35.3231 &  34.1 & --0.882 & 0.051 & --1.038 & 0.127 & \dots \\
    218.5853 & 32.8440 &   0.4 & 218.5850 & 32.8439 &   1.8 &    \dots &   \dots & \dots & --0.708 & 0.151 & >--1.17 & \dots & \dots \\
    218.7514 & 33.7390 &  14.0 & 218.7514 & 33.7389 &  95.7 & 218.7517 & 33.7392 & 240.6 & --0.870 & 0.032 & --0.917 & 0.073 & \dots \\
       \dots &   \dots & \dots & 215.3682 & 34.2739 &  14.6 &    \dots &   \dots & \dots & <--1.68 & \dots &  >0.01 & \dots & \dots \\
    220.7384 & 35.2715 &   0.3 &    \dots &   \dots & \dots &    \dots &   \dots & \dots & >-0.56 & \dots &   \dots & \dots & \dots \\
\hline
\end{tabular}
\tablefoot{Extract of the cross-match table. Columns (1--9) represent RA, Dec and total flux density for C4, C5 and C6 catalogs respectively; Columns (10--13) give spectral index and its error for Apertif-HBA and HBA-LBA frequencies respectively. When the total flux density measurement is missing in one of the catalogs, the spectral index limit is calculated using 5 times the local RMS as an upper limit on the source flux density at this frequency. Column (14) -- redshift estimate of the corresponding HBA source. The full table containing 12756 entries will be available in machine-readable format through CDS.}
}
\end{table*}


\section{Spectral indices}
\label{sec:SI}

Taking advantage of the available images, we can perform a detailed spectral index analysis. 
Hereafter, the spectral index between two surveys is defined as $\alpha^{\nu_1}_{\nu_2}=\ln{(S_{\nu_1}/S_{\nu_2})}/\ln{(\nu_1/\nu_2)}$, where $S_{\nu_1}$ and $S_{\nu_2}$ are the total flux density measurements, and $\nu_1, \nu_2$ are the corresponding survey frequencies (see Table~\ref{tab:data}). Having three frequencies we also refer to $\alpha^{1400}_{150}$ and $\alpha^{150}_{54}$ as ``high'' and ``low'' frequency spectral index respectively, $\alpha_{\rm high}$ and  $\alpha_{\rm low}$. 

In Table~\ref{tab:tab_si} sources from the C4, C5 and C6 catalogs are listed. If a source at a given location is present in the corresponding catalog, its coordinates and total flux density are given. When two flux density measurements are available, the spectral index is given along with its error. When a source is missing in one of the catalogs, a spectral index limit is  estimated using 5 times the local RMS of the image with the missing source. In the last column we give the redshift estimate if available (see Sect.~\ref{sec:SI_z}). We note that the three images are not contemporaneous and  some of the extreme apparent spectral indices can therefore reflect an intrinsic variability of the sources.

Below, in Sects~\ref{sec:SI-distribution} through \ref{sec:SI_z}, we discuss the spectral properties of the sources from Table~\ref{tab:tab_si}. We also identify extended radio sources with interesting spectral index distributions and build their spectral index maps in Sect.~\ref{sec:si_maps}. 


\subsection{The distribution at low and high frequencies}
\label{sec:SI-distribution}

\begin{figure}
    \centering
    \includegraphics[width=\columnwidth]{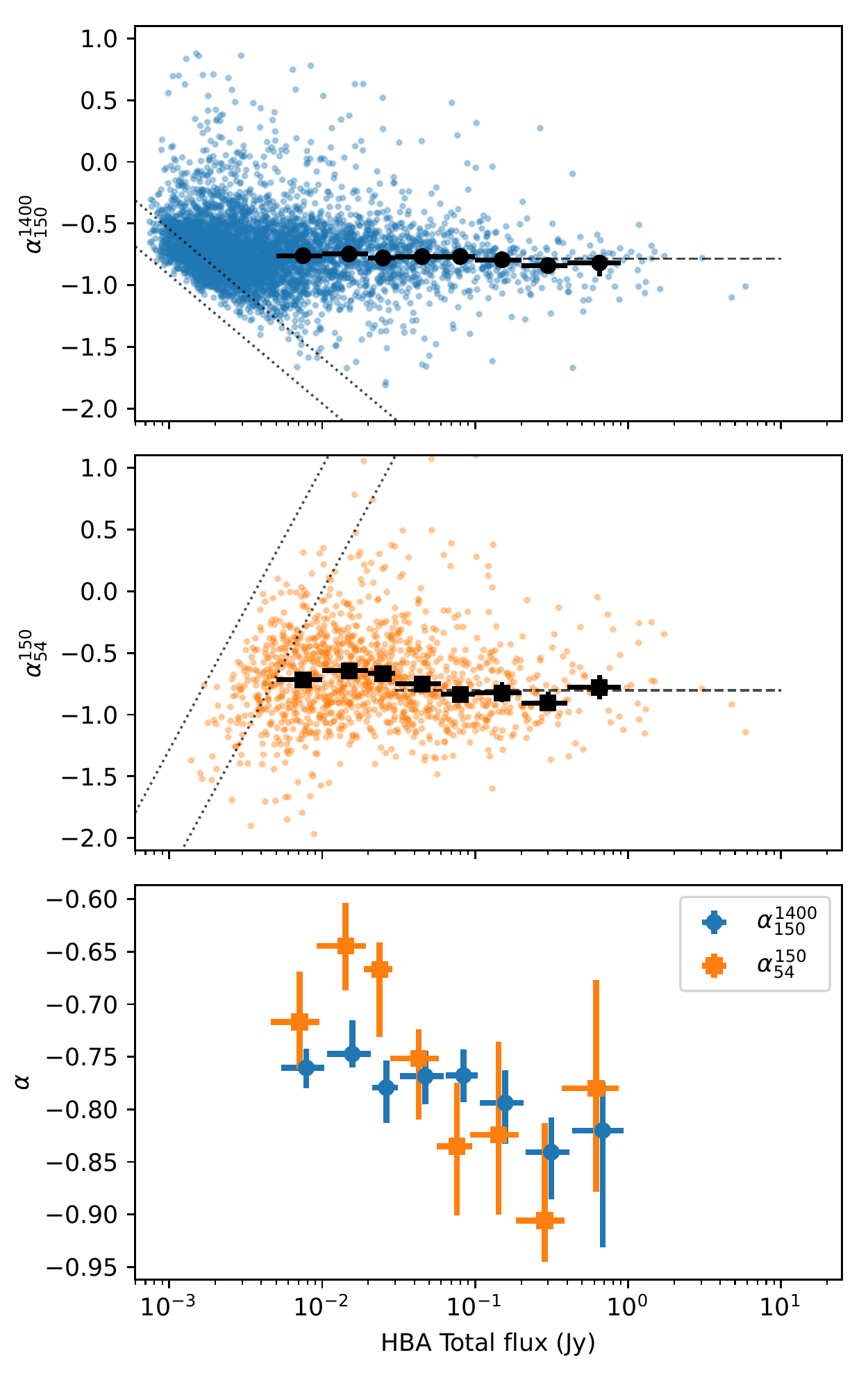}
    \caption{Spectral indices measured between 1400 and 150~MHz (top) and 150 and 54~MHz (middle) versus the 150~MHz HBA total flux density. Dotted lines define spectral index limits calculated using minimal total flux density and completeness level of C4 and C6. The median values calculated for the sources above the completeness limits (--0.77 and --0.73) are shown with horizontal dashed lines. Markers with bars show median spectral index calculated inside 8 flux density bins (also shown separately in the bottom panel). The bins are shown with horizontal bars. The 90\% confidence intervals of the median values derived using bootstrap are shown with vertical bars. }
    \label{fig:si_flux}
\end{figure}

The distributions of the spectral indices $\alpha^{\rm 1400}_{\rm 150}$ and $\alpha^{\rm 150}_{\rm 54}$ for all sources are shown in Fig.~\ref{fig:si_flux} plotted as a function of the 150~MHz total flux density. 
The relatively better depth of the 150~MHz HBA survey means that weaker HBA sources can be detected in LBA and/or Apertif only if they have a suitable spectral index. We take these limits into consideration in the analysis presented in Sect.\ \ref{sec:color-color} and \ref{sec:si_maps} and for the interpretation of the results.
In Fig.~\ref{fig:si_flux} the dotted lines show the spectral index calculated for the C5 flux density and the completeness levels of C4 and C6 respectively (0.3 and 10\,mJy), separating the regions where the distributions are incomplete due to the sensitivity limits of the Apertif and LBA surveys. Note that the completeness levels of the smoothed catalogs C4 and C6 are only slightly different from the ones of the original ones C1 and C3 (see noise values in Table~\ref{tab:data}).

The median values of $\alpha^{\rm 1400}_{\rm 150}$ and $\alpha^{\rm 150}_{\rm 54}$ calculated for the sources with HBA total flux density above \mjy{30}, so not biased due to sensitivity limits, are $\alpha_{\rm high}=-0.79 \pm 0.01$ and $\alpha_{\rm low}=-0.80 \pm 0.02$ respectively (the errors here define a 90\% confidence interval calculated using bootstrapping). This is consistent with the results obtained in K22 (see discussion and references therein for more details). The major scatter of the spectral index comes from intrinsic properties of the sources while the median value remains robust. The distribution of the low frequency spectral indices is similar to one obtained for the original catalogs C2 and C3 by \cite{2021A&A...655A..40W}. These authors report a trend of spectral flattening towards low flux density which they explain by the increasing relative number of core-dominated compact AGN and star forming galaxies (SFGs). 

We binned the sources using their HBA total flux density into 8 intervals (bins) and calculated the median spectral index in each bin. For every median value we calculated the 90\% confidence interval using a bootstrap approach. This is needed because the distributions of spectral index inside a bin is non-Gaussian, especially at lower flux density where the completeness plays a role. The bin widths, median values of spectral index and the corresponding 90\% confidence intervals are shown in Fig.~\ref{fig:si_flux} with error bars. 

There is, indeed, a trend seen of spectral flattening down to 10\,mJy, in agreement within the errors with previous studies~\citep[e.g.,][]{2011A&A...535A..38I, 2016MNRAS.463.2997M, 2018MNRAS.474.5008D, 2021A&A...655A..40W}. The low-frequency spectral index $\alpha^{\rm 150}_{\rm 54}$ becomes significantly flatter than $\alpha^{\rm 1400}_{\rm 150}$ for sources with HBA total flux density below 30\,mJy (obviously, the leftmost bin in Fig.~\ref{fig:si_flux} is affected by the incompleteness of the LBA catalog C6). At the same time, the high-frequency spectral index remains below --0.75 for sources in all flux density bins. The spectral index over the full frequency span,  $\alpha^{\rm 1400}_{\rm 54}$, takes intermediate values between the two. 
Significant flattening of the median spectral index at lower frequencies can be explained by increasing fraction of sources with concave radio spectra having a peak around the lowest frequency (54\,MHz). This scenario is also supported by analyzing the dependency of the spectral index on redshift (Sect.\ref{sec:SI_z}).

\begin{figure*}
    \centering
    \includegraphics[width=\columnwidth]{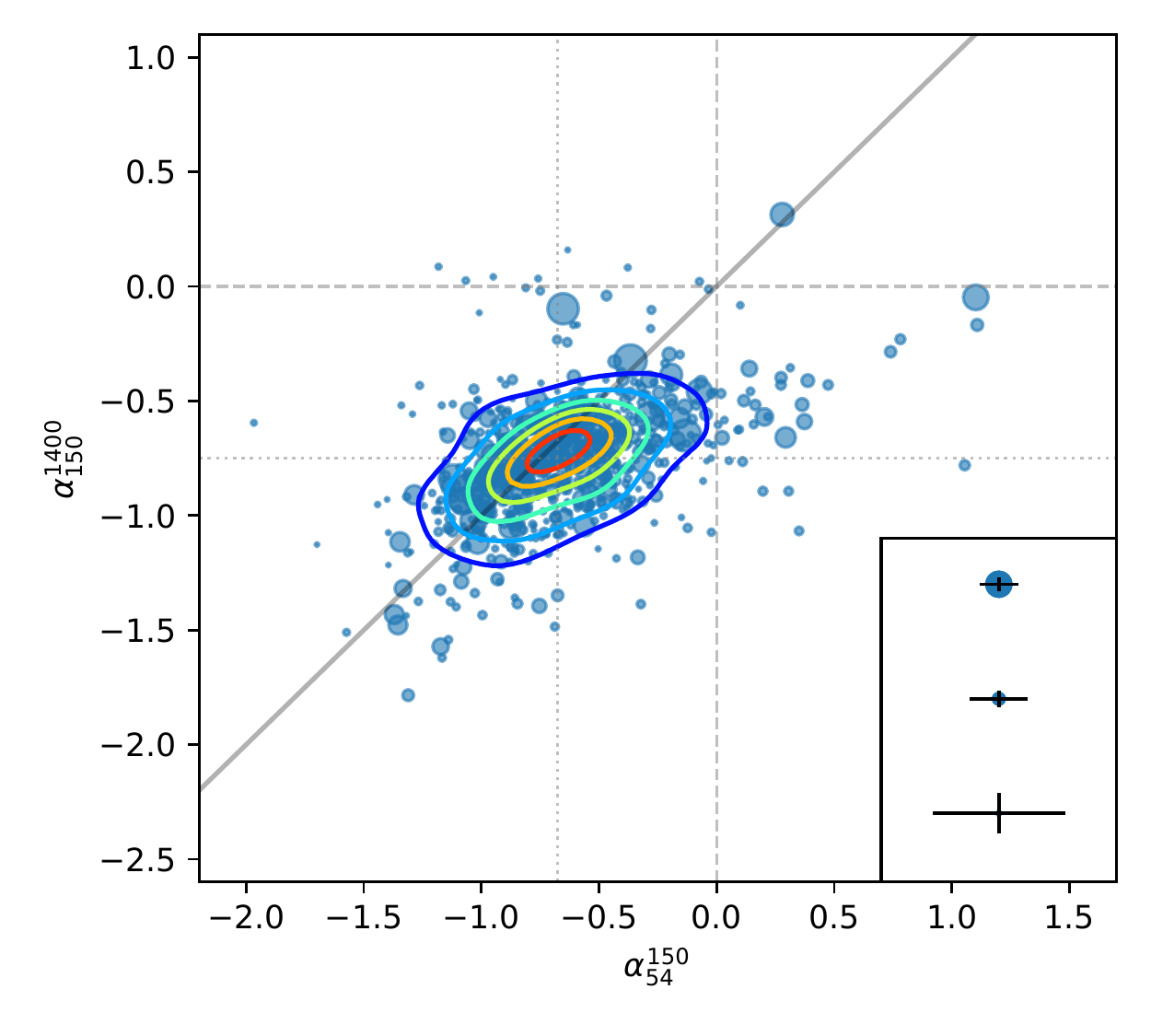}
    \includegraphics[width=\columnwidth]{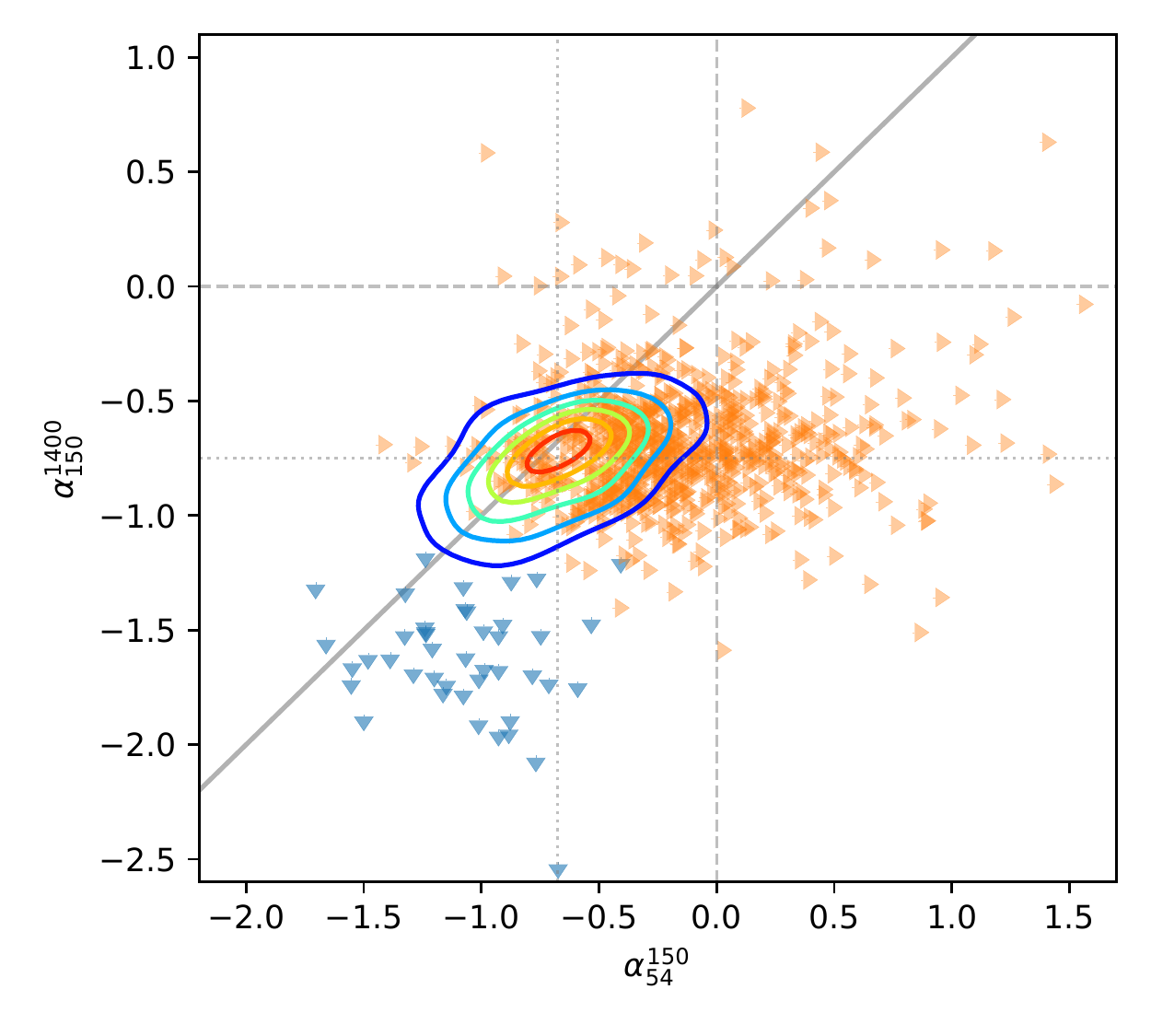}
    \caption{Spectral index diagrams of all sources, (see text for details) left panel shows sources detected in catalogs C4, C5 and C6. Size of the markers is proportional to source S/N in C5. The typical errors are shown in the bottom-right corner. The right panel shows the spectral index limits of the sources missing in either C4 or C6 estimated from noise RMS of the images M4 and M6 at the location of a source. The contours show Gaussian-kernel density estimate. The dashed lines indicate zero spectral index, while the dotted lines show median values of $\alpha^{\rm 1400}_{\rm 150}$ and $\alpha^{\rm 150}_{\rm 54}$ calculated for the common sources. 
    }
    \label{fig:si_si}
\end{figure*}


\subsection{Color-color diagram}
\label{sec:color-color}

The availability of the data at three frequencies allows us to expand the analysis of the spectral index by building a color-color diagram of $\alpha^{\rm 1400}_{\rm 150}$ vs $\alpha^{\rm 150}_{\rm 54}$ calculated as described above. The color-color diagram allow us to identify groups of sources with extreme spectral indices or that are deviating from a single power-law in the frequency range between LBA and Apertif frequencies. This in turn can tell us about the physical conditions of the sources. 

The color-color plot is shown in Fig.~\ref{fig:si_si}. The left panel shows the sources detected in all three catalogs C4, C5 and C6. The size of the markers is proportional to the S/N of a source in C5. The contours show the Gaussian-kernel density estimate. To minimize number of artifacts and outliers due to a complex source structure, we require a source to have be fitted by a single gaussian (\texttt{S\_Code='S'}) and have $S/N>15$ in the HBA catalog C5. 
Although a large fraction of the sources is located on the 1:1 line, indicating that they are characterized by a  power law spectral distribution without major breaks, Fig.~\ref{fig:si_si} shows that a significant fraction of the sources have spectral indices deviating from this line,  with the points being located below this line.
This is particularly clear for the group of sources with flatter spectrum ($-0.5<\alpha^{\rm 150}_{\rm 54}<1$) at low frequencies showing instead a steeper spectrum $-1<\alpha^{\rm 1400}_{\rm 150}<0.3$) at higher frequencies. A similar trend has been also reported by Boehme et al.\ (in prep.), who cross-matched LOFAR data with other surveys. 

To further investigate this trend, we have expanded the color-color plot by deriving also the limits to the spectral indices. If a source in C5 (HBA) has no counterpart in C4 (Apertif) or C6 (LBA), we estimated the limits of its spectral indices using the RMS noise  of the corresponding images M4 and M6. The flux density limits were estimated using 5 times the noise  at the location of the corresponding source in C5. These limits are shown in the right panel of Fig.~\ref{fig:si_si}. This addition to the original color-color diagram is useful for confirming the trend of sources with flatter spectral indices at low frequencies (Fig.~\ref{fig:si_si}, right) and complementing the distribution of the sources detected by all three surveys. 

Thus, adding the limits has further confirmed the presence of  a group of sources showing a flattening at low frequency. Peaked spectrum sources, defined as having an inverted spectral index at low frequency ($\alpha_{\rm low} > 0$ and a steep spectrum at high frequencies, are also found. 
Spectrum flattening or turnover at lower frequencies is also seen in Fig.~\ref{fig:si_flux} and might be a manifestation of absorption processes like synchrotron self-absorption or free-free absorption. The corresponding sources can be both compact synchrotron sources and/or star forming galaxies~\citep{2017ApJ...836..174C, 2018MNRAS.474..779G, 2021A&ARv..29....3O}. 

Another interesting group of sources in the color-color plot are the USS, with spectral index steeper than $-1.2$, either at low or high (or both) frequencies. 
The depth of the Apertif image, although not enough to directly detect many USS, allows us to put tight limits to identify a larger group of them. 
Considering the Apertif flux limit of \ujybeam{200} ($5\sigma$ noise  of image M4) for reliable source detection, the counterpart source with USS in image M5 should have a peak flux density $\gtrsim$\mjybeam{2.8}. There are 2497 HBA sources with peak flux density above this limit. To avoid artifacts and sources with a complex structure, we further require them to be fitted with a single gaussian (S\_Code="S") and have $S/N>15$ in C4 catalog resulting in  1743 sources. With these restrictions, the fraction of USS sources including the limits from Apertif non-detections, is about 6--9\%.

A number of sources (110) have USS also at low frequencies,  between 150 and 54 MHz. The nature of these sources is interesting to understand, as they can be particularly old remnant sources, or high redshift sources. Thirty-eight of these sources have redshift estimates (see Sect.~\ref{sec:SI_z}) with a maximum redshift of 5.22 and median value of 1.37, suggesting that some of these USS sources could be at high $z$ (but see Sect.\ \ref{sec:SI_z}). We note that a more careful redshift association might be needed to study individual sources.
 
At the same time, some of the USS sources  at low frequencies are candidate remnants. This is illustrated by the fact that we find cases of sources with extended USS emission  associated with low redshift galaxies. For example, a diffuse source J142957+325516 has a low frequency spectral index $\alpha_{\rm low} =-1.8$ and an estimated redshift of $z=0.24$. Other interesting examples of sources with extended USS emission are J143623+353430 at $z=0.74$ and J143134+332321 at $z=1.3$. These sources show properties characteristic of remnant radio sources, as discussed by \cite{2017A&A...606A..98B, 2021A&A...653A.110J}. 
 
Other cases of extended objects where only part of the  emission is USS will be discussed in Sect.~\ref{sec:si_maps} for sources larger than \amin{1}.
In these sources, the remnant emission appears to be co-existing with a new phase of activity, suggesting they are examples of restarted sources.

We also found sources in the Apertif image which do not have a counterpart in the LOFAR images. These sources have very inverted spectra. An example is J144334.8+334012 
which has an Apertif total flux density of 14 mJy, but has an upper limit of 0.35 mJy in the LOFAR HBA image, giving a lower limit of $\alpha^{1400}_{150} > 1.65$. This source was also detected in the Green Bank Telescope (GBT) survey at 4.85 GHz with a flux density of 34 mJy~\citep{1991ApJS...75.1011G}.


\subsection{Spectral index and redshifts}
\label{sec:SI_z}

As mentioned in the Introduction, a number of early studies have reported a correlation between spectral index and redshift and providing several possible explanations for this. However,  there are also works where no correlation was found~\cite[e.g.,][]{2017MNRAS.469.3468C}.
It is interesting to probe this dependency using our data with redshifts estimates available for a large number of sources.

We used the redshift estimate labels as `best' (\texttt{zbest}, spectroscopic or photometric) from the published value-added \bootes~HBA catalog~\citep{2019A&A...622A...3D, 2021A&A...648A...3K}. We cross-matched this catalog with C5 with a \asec{5} matching radius, resulting in more than 3000 redshift associations. These redshift estimates are given in the last column of Table~\ref{tab:tab_si}. We note that due to the relatively low resolution of C5, there might be some cross identification mistakes, and one should be using individual redshift estimates with care. 

\begin{figure}
    \centering
    \includegraphics[width=\columnwidth]{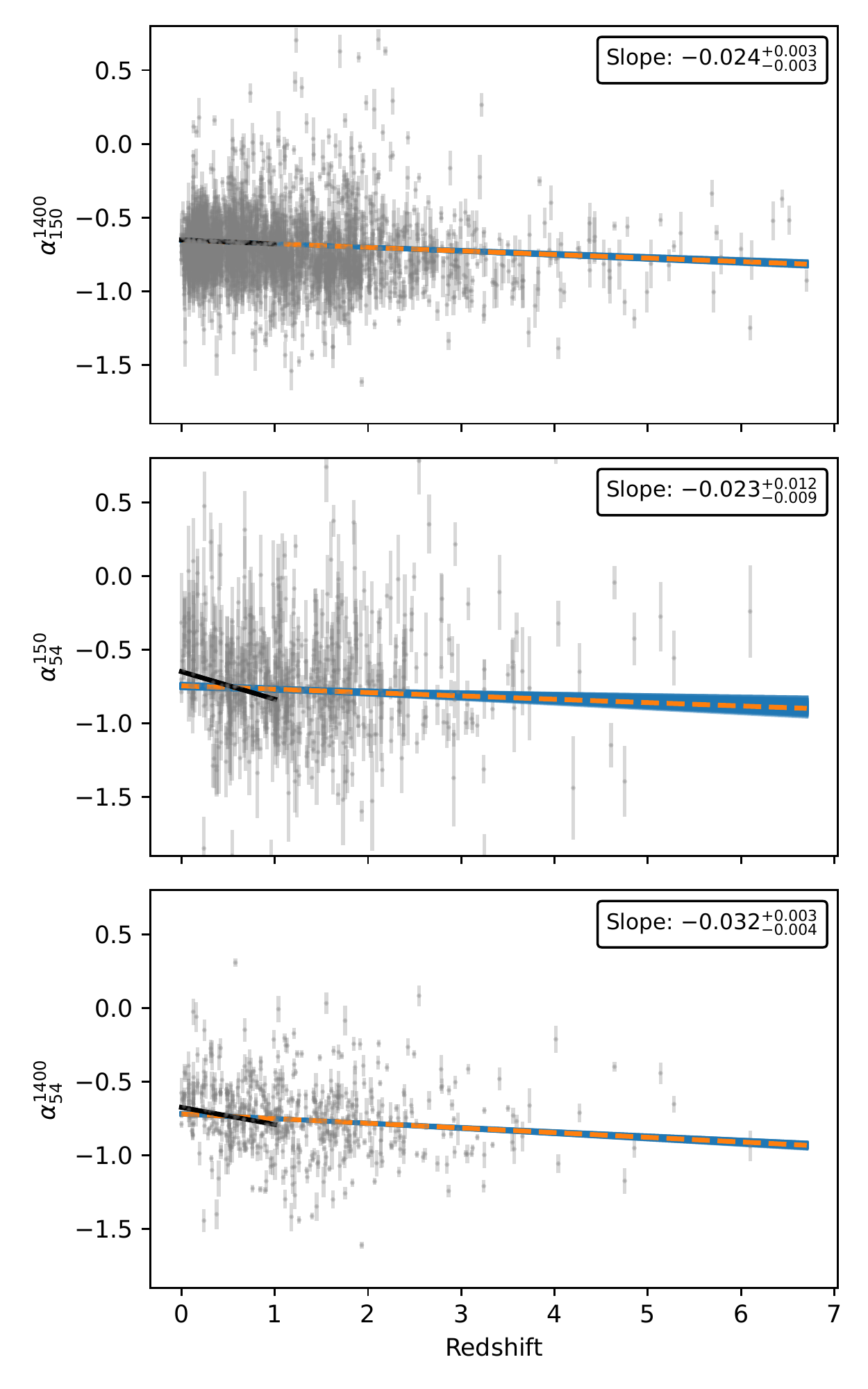}
    \caption{Spectral index as a function of redshift. Solid lines show random samples drawn from MCMC posterior distribution, dashed lines show linear model with MAP estimate. The corresponding slope coefficient is shown with its 95\% confidence interval in upper right corner. The same linear fit for $z<1$ is shown with solid black lines. The parameters are given in Table~\ref{tab:parameters}.}
    \label{fig:si_redshift}
\end{figure}


The spectral index  is plotted against redshift in Fig.~\ref{fig:si_redshift} separately for Apertif-HBA, HBA-LBA and Apertif-LBA frequency pairs. 
We estimated the probabilistic parameter distribution of a linear model fitted to the data with a Markov Chain Monte Carlo (MCMC) approach using the \texttt{emcee} package\footnote{https://emcee.readthedocs.io}~\citep{2013PASP..125..306F}. The model predictions for random samples from posterior parameters distribution are drawn along with the maximum a posteriori (MAP) estimate. The corresponding linear model coefficient is shown in the upper right corner of each panel with its 95\% confidence interval. 

There is a trend of spectral steepening with redshift seen in all three data sets. It can be seen that the redshift dependency is different for the low frequency spectral index and the high frequency one. The former drops steeply at lower redshifts and shows a slower decrease at higher redshifts, while the latter shows only a weak linear trend with a slope of $\sim 0.03$ over the entire redshift range. To illustrate this, and to compare with  previous studies, we exclude objects with redshift larger than  1 from the sample and repeat the analysis. The obtained model parameters are listed in Table~\ref{tab:alpha_redshift}. The slope coefficients found for the last two models are similar to those reported by~\cite{2018MNRAS.480.2726M} for this redshift range ($0<z<1$). 

\begin{table}[]
    \centering
    \caption{Parameters of linear model fit $\alpha(z) = az+b$ \label{tab:alpha_redshift}}
{\renewcommand{\arraystretch}{1.4}   \begin{tabular}{ccc}
    \hline\hline
        $\alpha$ & $a$ & $b$ \\
    \hline
      $\alpha^{1400}_{150}$ & $-0.024^{+0.003}_{-0.003}$ & $-0.655^{+0.005}_{-0.004}$ \\
      $\alpha^{150}_{54}$ &  $-0.023^{+0.012}_{-0.009}$ & $-0.746^{+0.014}_{-0.018}$ \\
      $\alpha^{1400}_{54}$ & $-0.032^{+0.003}_{-0.004}$ & $-0.719^{+0.006}_{-0.005}$ \\
      $\alpha^{1400}_{150}$ & $-0.028^{+0.013}_{-0.013}$ & $-0.652^{+0.008}_{-0.008}$ \\
      $\alpha^{150}_{54}$ & $-0.187^{+0.052}_{-0.052}$ & $-0.650^{+0.031}_{-0.030}$ \\
      $\alpha^{1400}_{54}$ & $-0.116^{+0.018}_{-0.015}$ & $-0.675^{+0.009}_{-0.011}$ \\
    \hline
    \end{tabular}}
    \tablefoot{First three rows correspond to the full redshift range, while the last three rows show the fit parameters for objects with $z<1$. The errors indicate a 95\% confidence interval for each parameter.}
    \label{tab:parameters}
\end{table}

A steeper redshift dependency for the low-frequency spectral index can be naturally explained if most  sources have a peaked spectrum and the intrinsic spectral turnover frequency is close to the LBA one (54 MHz). In this case, the LBA flux density of a source is measured near the spectrum peak and the resulting spectral index between the LBA and another frequency  strongly depends on  redshift. At the same time, the high-frequency spectral index does not trace the spectral turnover in most cases, and its redshift dependency remains weak. 
The overall common  trend with a slope of $\sim$0.03 seen in all three samples can be attributed to spectral steepening due to more intense inverse Compton losses at higher redshifts or to a simple spectral index -- luminosity correlation~\citep[see][for references]{2018MNRAS.480.2726M}. 

We note that the sample considered here represents a complex mixture of sources with different types of spectra. For example, in the the Apertif-HBA cross-match sample there are more sources with flat or inverted spectra than in the HBA-LBA sample, resulting in a different spectral index behavior between these groups. An increasing number of star forming galaxies appearing at low redshifts might also play a role in the observed spectral flattening. Finally, the paucity of LBA-based spectral indices smaller than --0.8 seen at low redshifts in Fig.\ \ref{fig:si_redshift} could in principle be due to some systematic flux scale offset for the LBA, but this is to be investigated outside of this work. To better understand the nature of the spectral index-redshift correlation, a separate detailed study is required which takes into consideration the properties of various source populations.


\subsection{Spectral index structure of selected extended sources}
\label{sec:si_maps}

The analysis presented above is based on the integrated flux density of the sources detected by PyBDSF.
However, thanks to the good angular resolution of the available images, many extended sources can be seen  and for them the resolved spectral index images can be derived.  
Thus, one of the new exciting possibilities provided by joint Apertif and LOFAR surveys is to explore the structure of the spectral index within a source and connect this to the evolution of the radio source.  The availability of three frequencies makes this analysis more challenging, but also more rewarding, allowing to expand on what has already been presented for the Lockman Hole ~\citep{2021Galax...9...88M, 2021A&A...648A...9M}.

For this, we have selected sources with a deconvolved size larger than \amin{1}. 
This limit was chosen to be a few restoring beam sizes large to make sure the spectral index structures are well resolved. 
In total, seventy-four sources larger than \amin{1} were found. The resulting sample is not large enough for a statistical study, but it allows us to investigate the spectral index structure in detail at least for some sources. For the purpose of this paper we have carried out most of the analysis/characterization of these resolved spectral indices by visual inspection.

Using images M4, M5 and M6, we constructed two spectral index maps of $\alpha^{\rm 1400}_{\rm 150}$ and $\alpha^{\rm 150}_{\rm 54}$ respectively. 
Spectral index detections were derived for pixels with signal at both frequencies above 3 $\sigma$ of the local noise. 
Instead, if at one frequency (typically LBA or Apertif) no detection was available, the limit was estimated by replacing the value of the pixels with the value of 3 $\sigma$ of the local noise. If both images are below 3 $\sigma$ then the pixels were replaced with NaN values. 

\begin{figure*}
    \centering
    \includegraphics[width=16cm]{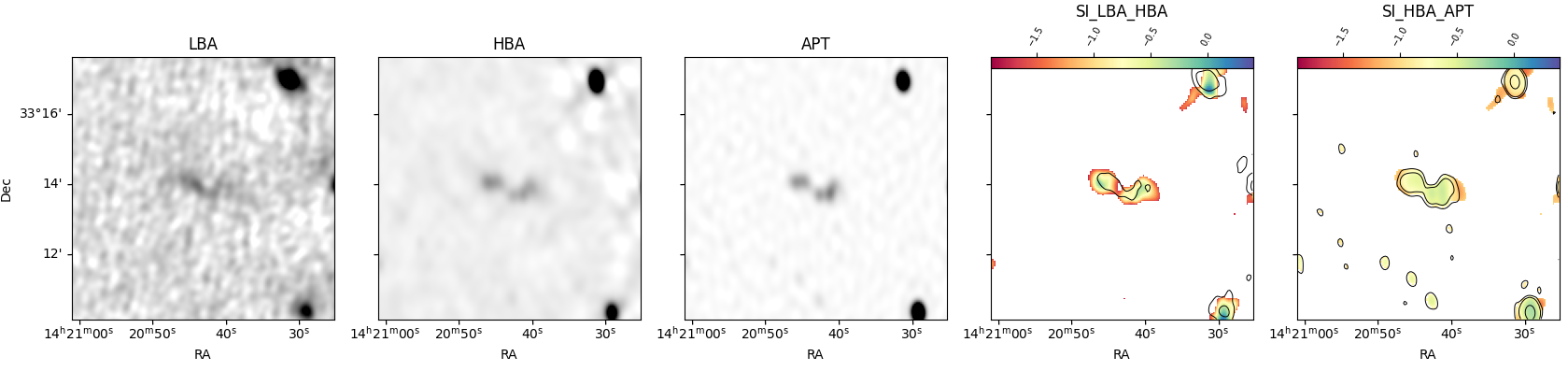} \\ 
     \includegraphics[width=16cm]{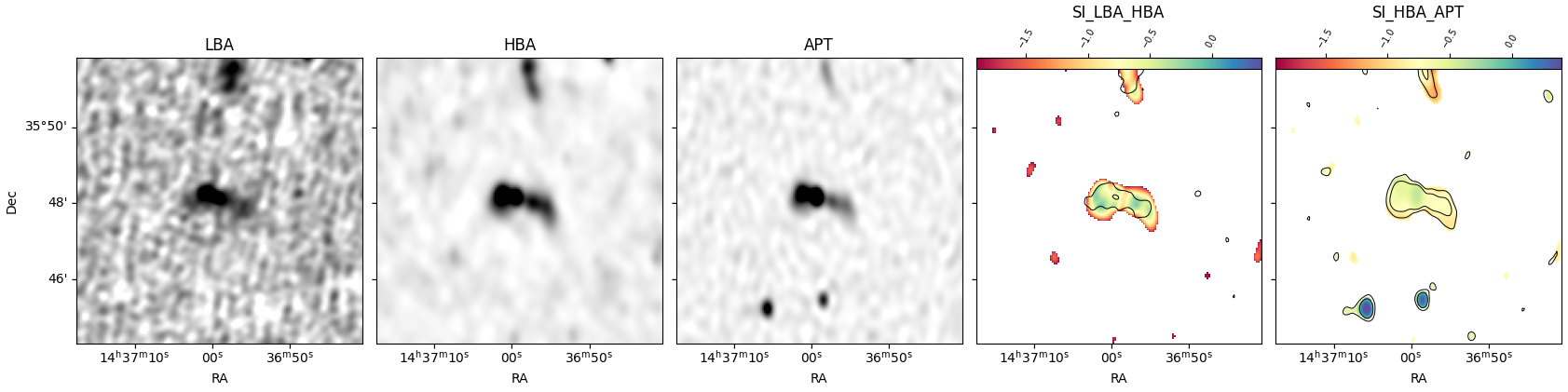} 
    \includegraphics[width=16cm]{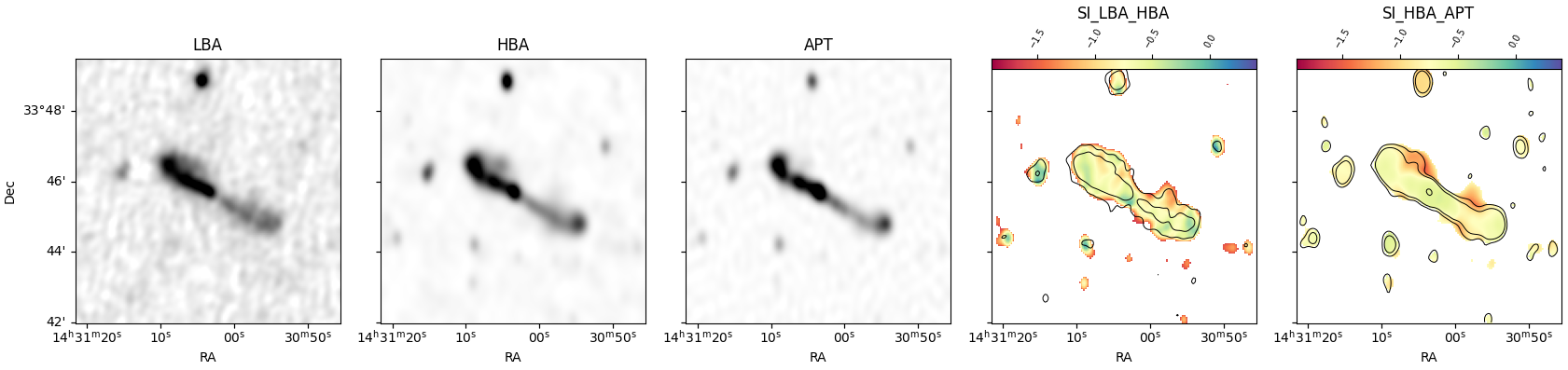} \\        
    \includegraphics[width=16cm]{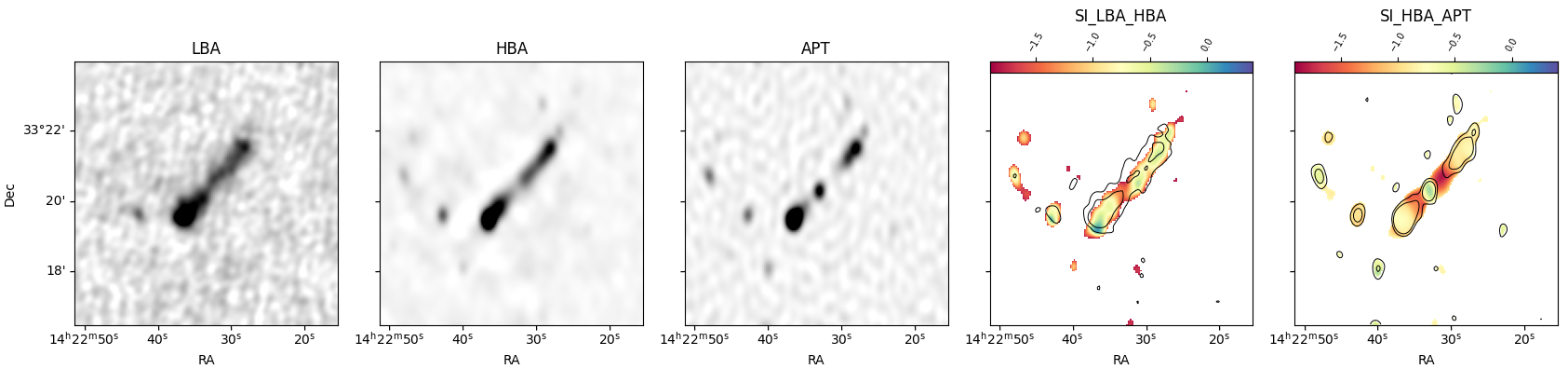} \\
    \includegraphics[width=16cm]{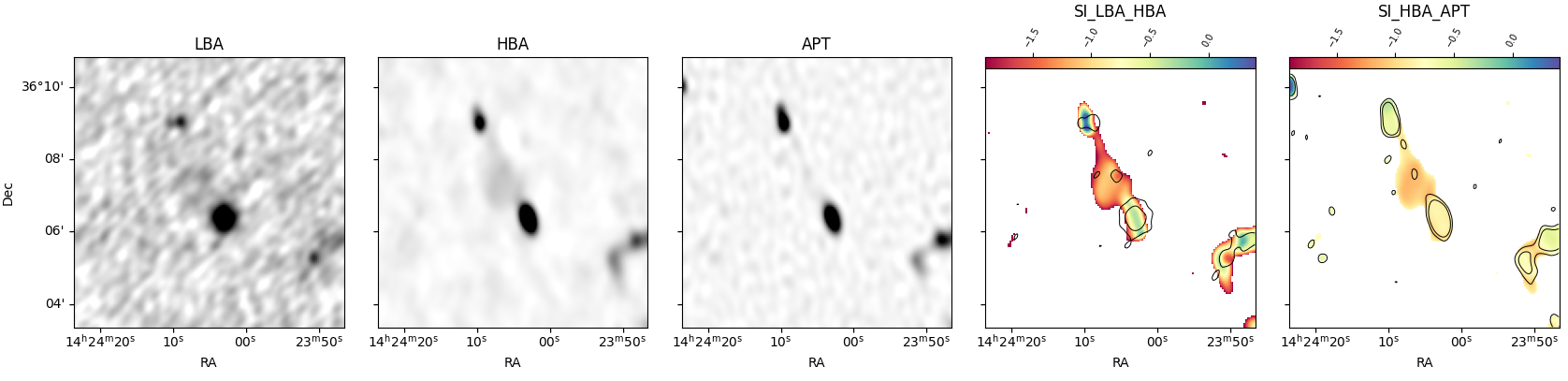} \\
 \includegraphics[width=16cm]{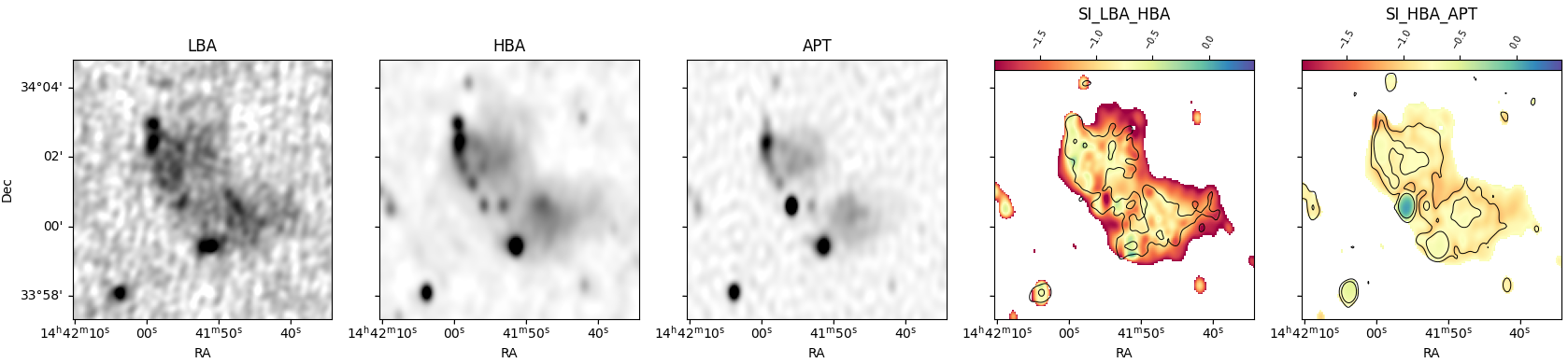} \\
    \caption{Examples of FRI (top two) and FRII (third and fourth) radio sources where the extended spectral index can be studied. See text for details. From left to right, LBA, HBA and Apertif images, Spectral indices LBA-HBA (with superposed the LBS contours) and HBA-Apertif (with superposed the Apertif contours). The contour levels are 3, 10, 150 $\sigma$, where $\sigma$ is the local noise level of LBA (left) and Apertif image (right, see text)}. 
    \label{fig:InterestingExamplesSI-1}
\end{figure*}

\begin{figure*}
    \centering
    \includegraphics[width=18cm]
{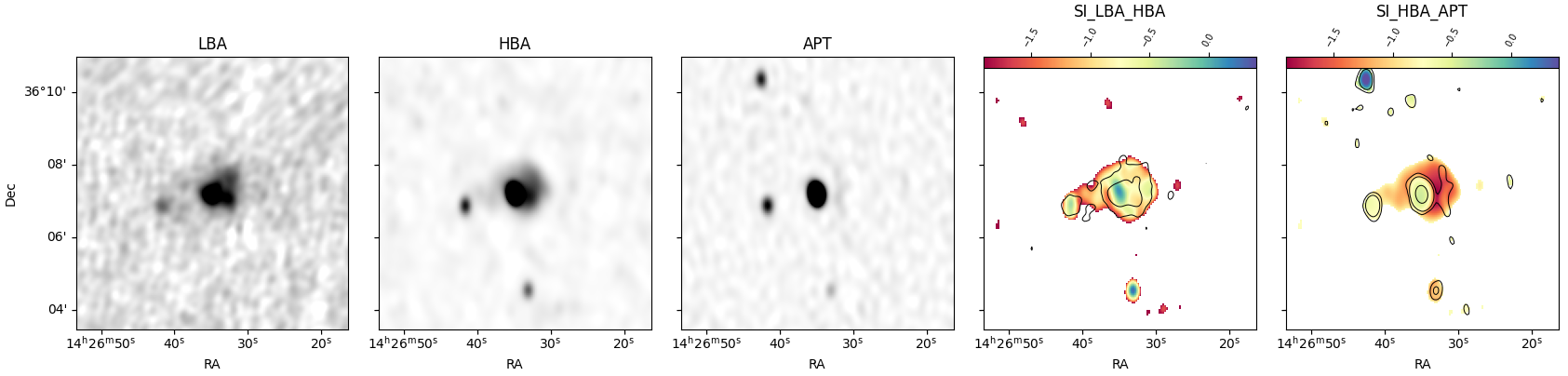} \\
      \includegraphics[width=18cm]
{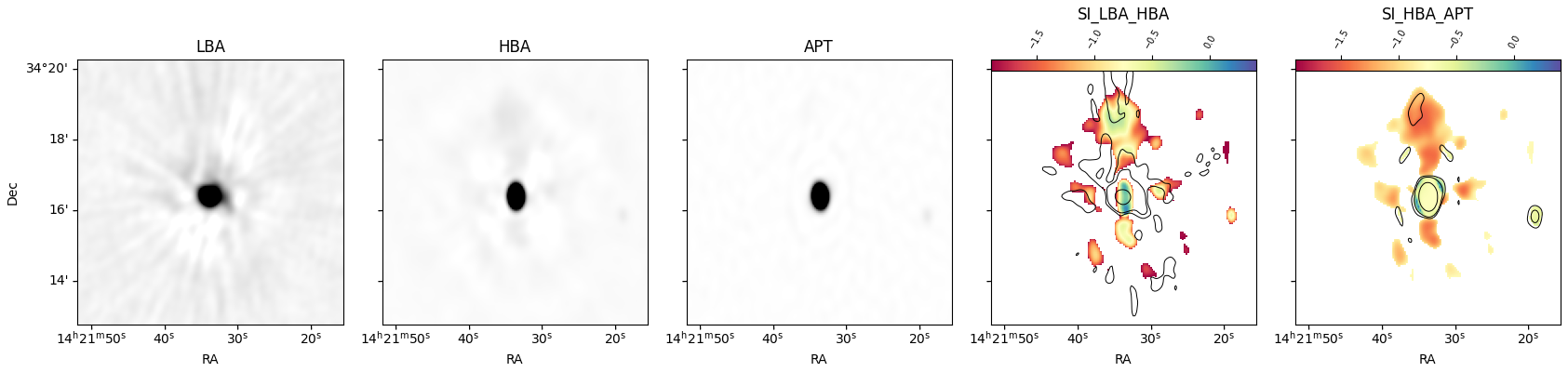} \\
    \includegraphics[width=18cm]
{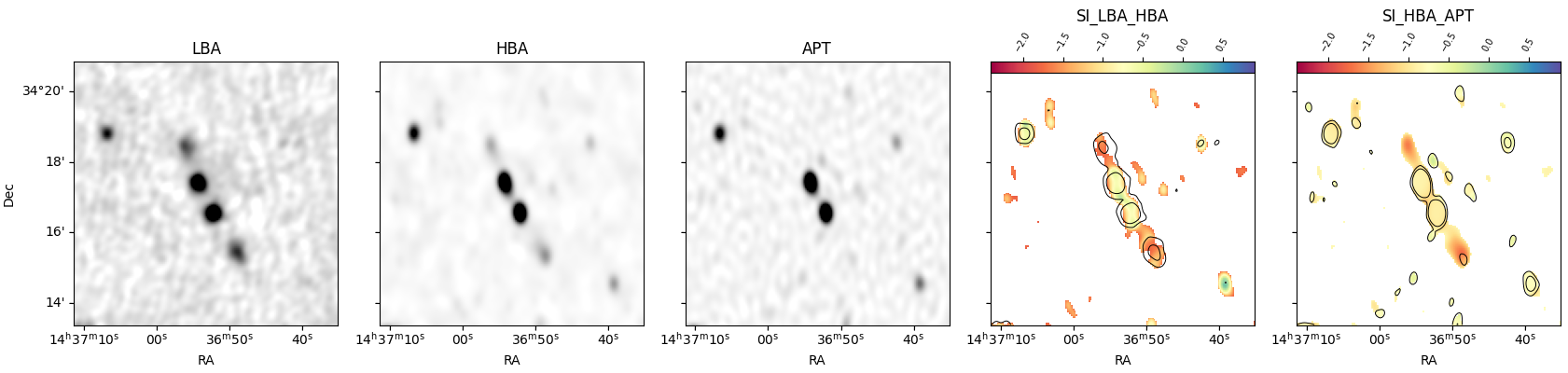} \\
  \includegraphics[width=18cm]
{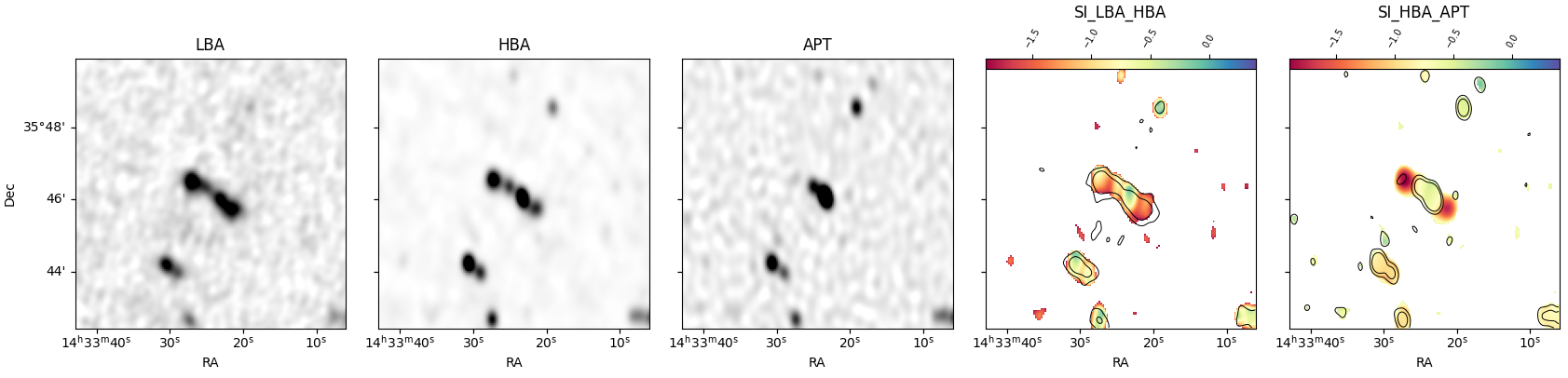} \\
   \includegraphics[width=18cm]
{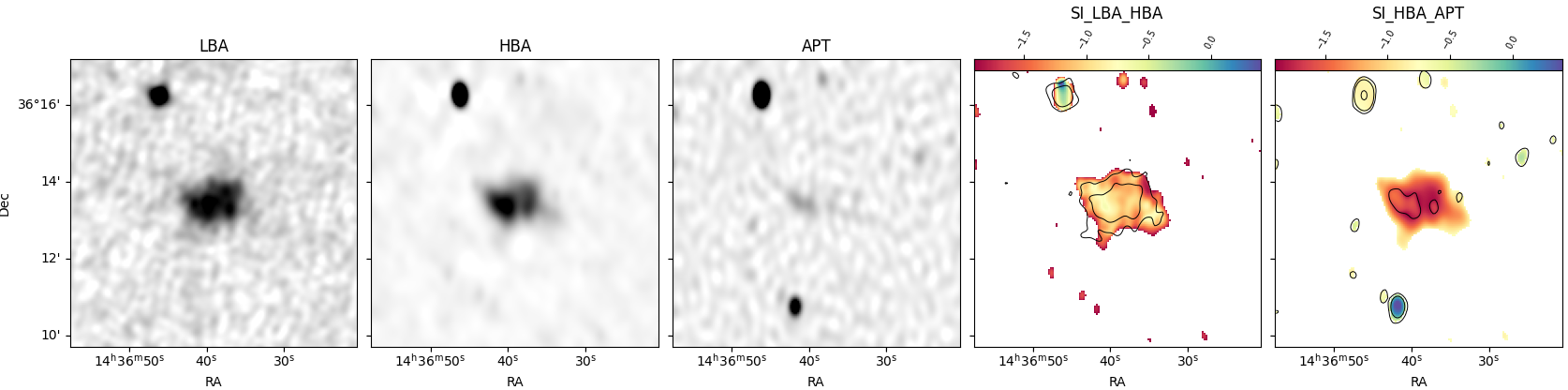} \\
    \caption{Examples of sources with USS regions. Images and contours as in Fig. \ref{fig:InterestingExamplesSI-1}. The top two show an active, inverted spectrum core and a USS extended region. The third and fourth are double-double type sources, where the outer is USS. All these sources can be  considered as restarted. The last source is an example of remnant source, with the entire emission USS.
    }
    \label{fig:InterestingExamplesSI-2}
\end{figure*}


The 74 selected sources show a large variety of spectral index structures. In Figs \ref{fig:InterestingExamplesSI-1} and \ref{fig:InterestingExamplesSI-2} we illustrate some of the more interesting cases. In these figures we present (from left to right) the intensity  LBA, HBA and Apertif images, while the last two images show the spectral index distribution. The LBA-HBA spectral indices have overlaid the LBA total flux density contours: outside these contours the spectral index represent a lower limit. Empty contours and white color indicate regions where both LBA and HBA emission is below 3$\sigma$. The rightmost images show the high-frequency spectral index HBA-Apertif with superposed contours of the Apertif  flux density. Spectral indices outside the contours (i.e.\ regions without Apertif detection) should be considered as upper limits. 

In Fig.\ \ref{fig:InterestingExamplesSI-1} we show examples of the spectral index structure in FR~I and FR~II sources.  The distribution follows what is expected for these sources and, in fact, it shows how the spectral index can help in classifying the sources even when the total intensity show a complex structure. 
The first two sources are FR~I and also the spectral indices indicate a relatively smooth distribution along the jets/lobes as expected for this type of sources.  
The morphology of the third source is more difficult to classify, having elements of both FR~I and II. The spectral index indicates the presence of a possible hot-spot in the south-west lobe, more clearly seen at low frequencies, while no clear hot-spot is seen in the north-east lobe.

Sources 4 and 5 in Fig.~\ref{fig:InterestingExamplesSI-1} are clear cases of FR~II sources, with flat-spectrum emission at the edges of the lobes (where the hot-spots are located) and steep spectrum emission in between, tracing the back-flow emission deposited as the jet proceeds (i.e.\ resulting in the typical trend of older plasma situated closer to the core). Interestingly, the fifth source shows that this steep spectrum emission (albeit not extreme, note that the spectral indices at low frequencies are lower limits) is extending perpendicular to the jet axis, perhaps indicating a change of its position angle in the past  of the source or an interaction with a companion, as suggested, for example, for 3C\,321, a well studied radio galaxy with a similar radio morphology~\citep{2008ApJ...675.1057E}. 
The last source at the bottom of Fig.\ \ref{fig:InterestingExamplesSI-1} is another complex case, suggesting an FR~II structure, but with flat-spectrum hot spots at the end of the jets seen only at low frequencies as well as a prominent core at high frequencies (which could also be a compact foreground source). The diffuse structure has a steep, but not extreme spectral index, and is suggestive of plasma tracing the motion of the source in a dense medium (perhaps a cluster or group). 

Thus, the combination of morphology and spectral index is adding more details to the classification based only on the morphology, making the separation of FR~I and II less sharp and more complex. Adding the information on the spectral index to the recent work by~\cite{2019MNRAS.488.2701M}, who probed this dichotomy using LOFAR observations, may further explain the  parameters influencing the radio morphology.

In addition to these sources, we have also found a number of cases where a large fraction (or even all) of the emission is USS, especially at high frequencies.
As already discussed at the beginning of  Sect.\ref{sec:SI}, these spectral properties are considered signatures of remnant emission of a previous phase of activity which has now stopped. In a number of cases we find the presence of an active region co-existing with the remnant one, confirming what was found in the study of the Lockman Hole using LOFAR HBA and Apertif images and suggesting that a dying phase has been followed by restarted activity. These radio sources are particularly interesting for understanding of the evolution of radio galaxies.

In Fig.\ \ref{fig:InterestingExamplesSI-2} some examples of these cases are shown. The top two sources show the presence of active cores (which can be identified by flat or inverted spectra) while the extended emission around them is USS at high frequencies (with most of the HBA-Apertif spectral index being an upper limit).  Thus, the new phase of activity indicated by an active core appears to have started before the remnant emission has had time to disappear. 

Sources 3 and 4 in Fig.~\ref{fig:InterestingExamplesSI-2}, and in particular source 3,  are  examples of a double-double-like structure, in other words sources with two sets of symmetric lobes likely resulting from two phases of activity~\citep[see e.g.,][and references therein]{2019A&A...622A..13M}, at least as they appear in the HBA image. Interestingly, the outer pair of lobes is not seen in the Apertif image, resulting in these lobes being characterized by an USS at high frequency (HBA-Apertif). 

According to the calculations presented in \cite{2021A&A...648A...9M}, USS at high frequencies (HBA-Apertif) imply ages of the remnants (i.e.\ the time passed since the last re-acceleration of the electrons in a magnetic field of order of $B_{\rm eq} = 3$ $\mu$G) to be in the range between 160 and 320 Myr. Within this period, a new phase of activity can start in such sources, as  observed in the Lockman Hole \citep{2021A&A...648A...9M}.

The last source in Fig.~\ref{fig:InterestingExamplesSI-2} has an amorphous structure and the entire emission is USS at high frequencies. We consider this a nice example of a remnant radio source. 
We do not find any extended source larger than \amin{1} where all emission is USS down to the lowest frequency. This may indicate that the remnant emission disappears below our detection limit for dying sources which are too old.  

\section{Conclusions}

We have presented an image of a 26.5 square degree region in the \bootes\ constellation obtained at 1.4~GHz using the Apertif system on the WSRT. This image
was made using  an improved calibration and imaging pipeline for Apertif. The main improvement offered by this new pipeline is that it incorporates direction-dependent calibration, allowing to correct the Apertif images for imaging artifacts due to faulty PAF elements and to produce  images of much higher quality compared to those of the first Apertif data release. With this pipeline we processed 187 Apertif data sets and released a mosaic image matching the \bootes\ deep field observed earlier by the LOFAR telescope at 150 and 54\,MHz. From the Apertif image, we compiled a source catalog containing \napt sources which is complete down to 0.3\,mJy level. 

We smoothed the Apertif image and the publicly available LOFAR images to the same angular resolution and made the matching catalogs to study spectral indices of the sources. The distribution of low- and high-frequency spectral indices of the common sources, combined with limits,  show that the majority of  sources demonstrate a break or curvature in their spectra at low frequency. 

We also investigated the dependency of the spectral indices on  redshift and found a negative correlation which can be explained as a K-correction of the spectra with a break or curvature at low frequencies. Using the color-color diagrams we identified and discussed various types of interesting sources, including remnants, young AGN and sources with restarted activity.

As shown in this work,  Apertif  sensitivity and angular resolution provide a unique synergy with the LOFAR surveys at lower frequencies. The joint data analysis has a great potential for many astrophysical studies.  Re-processing of Apertif data is ongoing and we will focus on  other LOFAR deep fields, such as the Elais-N1 and Lockman Hole area, for our future continuum releases.

\begin{acknowledgements}

This work makes use of data from the Apertif system installed at the Westerbork Synthesis Radio Telescope owned by ASTRON. ASTRON, the Netherlands Institute for Radio Astronomy, is an institute of the Dutch Science Organisation (De Nederlandse Organisatie voor Wetenschappelijk Onderzoek, NWO). Apertif was partly financed by the NWO Groot projects Apertif (175.010.2005.015) and Apropos (175.010.2009.012).
This research made use of \texttt{Python} programming language with its standard and external libraries/packages including \texttt{numpy}~\citep{numpy}, \texttt{scipy}~\citep{scipy}, \texttt{scikit-learn}~\citep{scikit-learn}, \texttt{matplotlib}~\citep{matplotlib}, \texttt{pandas}~\citep{pandas} etc.
This research made use of Astropy,\footnote{http://www.astropy.org} a community-developed core Python package for Astronomy \citep{astropy:2013, astropy:2018}. 
The \texttt{radio\_beam} and \texttt{reproject} python packages are used for manipulations with restoring beam and reprojecting/mosaicking of the images. 
This research has made use of "Aladin sky atlas" developed at CDS, Strasbourg Observatory, France~\citep{2000A&AS..143...33B, 2014ASPC..485..277B}. 
BA acknowledges funding from the German Science Foundation DFG, within the Collaborative Research Center SFB1491 ''Cosmic Interacting Matters - From Source to Signal''
KMH acknowledges financial support from the grant CEX2021-001131-S funded by MCIN/AEI/ 10.13039/501100011033, from the coordination of the participation in SKA-SPAIN, funded by the Ministry of Science and Innovation (MCIN) and from grant  PID2021-123930OB-C21 funded by MCIN/AEI/ 10.13039/501100011033, by “ERDF A way of making Europe” and by the "European Union". 
JMvdH acknowledges funding from the Europeaní Research Council under the European Union’s Seventh Framework Programme (FP/2007-2013)/ERC Grant Agreement No. 291531 (‘HIStoryNU’).
LCO acknowledges funding from the European Research Council under the European Union's Seventh Framework Programme (FP/2007-2013)/ERC Grant Agreement No. 617199.
JvL acknowledges funding from Vici research programme `ARGO' with project number 639.043.815, financed by the Dutch Research Council (NWO).
DV acknowledges support from the Netherlands eScience Center (NLeSC) under grant ASDI.15.406
\end{acknowledgements}

\bibliographystyle{aa}
\bibliography{main}

\end{document}